\documentclass[twocolumn]{aa}
\usepackage[varg]{txfonts}

\usepackage{graphicx}
\usepackage{tikz}
\usepackage[breaklinks=true]{hyperref}
\bibpunct{(}{)}{;}{a}{}{,} 

\newcommand{\thru}{\ensuremath{\text{--}}}

\newcommand{\um}{\ensuremath{\,\mathrm{\mu m}}}

\newcommand{\Angstrom}{\ensuremath{\,\text{\r{A}}}}

\newcommand{\polr}{\ensuremath{P_\mathrm{r}}}
\newcommand{\thr}{\ensuremath{\theta_\mathrm{r}}}

\newcommand{\Pmin}{\ensuremath{P_\mathrm{min}}}
\newcommand{\Pmax}{\ensuremath{P_\mathrm{max}}}
\newcommand{\amin}{\ensuremath{\alpha_\mathrm{min}}}
\newcommand{\amax}{\ensuremath{\alpha_\mathrm{max}}}
\newcommand{\ainv}{\ensuremath{\alpha_0}}
\newcommand{\pp}{\ensuremath{\,\mathrm{\%p}}}

\defcitealias{1990MNRAS.245...46G}{GG90}
\newcommand{\Alumina}{\citetalias{1990MNRAS.245...46G}}
\newcommand{\PaperII}{Paper II}

\begin{document}

\title{Quantitative grain size estimation on airless bodies \\ from the negative polarization branch}
\subtitle{I. Insights from experiments and lunar observations
\thanks{Part of the codes and data used for this publication are available via GitHub: \url{https://github.com/ysBach/BachYP_etal_CeresVesta_NHAO}.}
}

\author{
  Yoonsoo P. Bach\inst{1, 2, 3}
  \and Masateru Ishiguro\inst{2, 3}
  \and Jun Takahashi \inst{4}
  \and Jooyeon Geem\inst{2, 3}
  \and \\ Daisuke Kuroda \inst{5}
  \and Hiroyuki Naito \inst{6}
  \and Jungmi Kwon \inst{7}
}

\institute{
  Korea Astronomy and Space Science Institute (KASI), 776 Daedeok-daero, Yuseong-gu, Daejeon 34055, Republic of Korea
  \and
  Department of Physics and Astronomy, Seoul National University, Gwanak-ro 1, Gwanak-gu, Seoul 08826, Republic of Korea
  \and
  SNU Astronomy Research Center, Department of Physics and Astronomy, Seoul National University, Gwanak-ro 1, Gwanak-gu, Seoul 08826, Republic of Korea
  \and
  Center for Astronomy, University of Hyogo, 407-2 Nishigaichi, Sayo, Hyogo 679-5313, Japan
  \and
  Japan Spaceguard Association, Bisei Spaceguard Center 1716-3 Okura, Bisei, Ibara, Okayama 714-1411, Japan
  \and
  Nayoro Observatory, 157-1 Nisshin, Nayoro, Hokkaido 096-0066, Japan
  \and
  Department of Astronomy, Graduate School of Science, The University of Tokyo, 7-3-1 Hongo, Bunkyo-ku, Tokyo 113-0033, Japan
  \\ \email{ysbach93@gmail.com, ishiguro@snu.ac.kr}}

\date{Received ; accepted }

\abstract
{
This work explores characteristics of the negative polarization branch (NPB), which occurs in scattered light from rough surfaces, with particular focus on the effects of fine particles. Factors such as albedo, compression, roughness, and the refractive index are considered to determine their influence on the NPB. This study compiles experimental data and lunar observations to derive insights from a wide array of literature. Employing our proposed methodology, we estimate the representative grain sizes on the lunar surface to be $D \sim 1\thru 2\um$, with $D \lesssim 2\thru4\um$, consistent with observed grain size frequency distributions in laboratory settings for lunar fines. Considering Mars, we propose that the finest particles are likely lacking ($D\gg 10\um$), which matches previous estimations. This study highlights the potential of multiwavelength, particularly near-infrared, polarimetry for precisely gauging small particles on airless celestial bodies. The conclusions provided here extend to cross-validation with grain sizes derived from thermal modeling, asteroid taxonomic classification, and regolith evolution studies.
}


\titlerunning{NIR polarimetry of airless bodies}
\maketitle

\section{Introduction}\label{s:intro}
With current aspirations for space exploration, numerous airless objects have been thoroughly investigated and leverage advanced scientific approaches, such as orbiters, flybys, and landers. These explorations have facilitated the study of the geology, shape, dynamics, and consequences of impacts on these celestial remnants (e.g., the recent DART mission by \citealt{2023Natur.616..443D}). Despite extensive exploration, direct visual observation of decamicron-scale dust particles has not been performed since few images have been able to resolve such fine scales on the surface. However, the presence of these fine particles can impact scattering or emission processes, thereby modulating the observed light. As a result, indirect methods, such as thermal modeling \citep{2013Icar..223..479G,2022PSJ.....3...47M} and polarimetric observations \citep{1986MNRAS.218...75G,2017AJ....154..180I,2018NatCo...9.2486I,2019JKAS...52...71B}, have been employed to explore the existence of these fine particles. We focus on the latter approach and suggest a novel way to estimate grain sizes using multiwavelength polarimetry.

In reality, any solid surface is composed of an irregular spatial distribution of dielectric coefficients, that is, has a degree of disorder, which is often practically impossible to describe mathematically. Thus, as a rough approximation, the disorder is described by using familiar terms, such as grain size, reflectance/albedo, composition, surface roughness, and macro/microporosity.

Typically, in Fresnel reflection or Rayleigh scattering, the strongest electric field vector aligns perpendicularly to the scattering plane (the plane formed by the incident and emitted light rays). However, at small phase angles, the strongest electric vector is predominantly parallel to the scattering plane for rough surfaces \citep[the first report was by][]{Brewster1861}. Recognizing this behavior, \cite{1929PhDT.........9L} proposed a modified linear polarization degree for use in observational and experimental studies:
\begin{equation}\label{eq: polr}
  \polr \coloneqq P \cos (2 \thr) \approx \frac{I_\perp - I_\parallel}{I_\perp + I_\parallel}~,
\end{equation}
where $P$ represents the total linear polarization degree, $\thr$ is the position angle of the strongest electric vector relative to the normal vector of the scattering plane ($0/180^\circ$: normal to the scattering plane, $90/270^\circ$: within the scattering plane), and $I_\perp$ and $I_\parallel$ denote the measured intensities of scattered light along the directions perpendicular and parallel to the scattering plane, respectively.

The polarimetric phase curve (PPC), represented by $\polr(\alpha)$, is plotted as a function of the phase angle (light source-target-observer angle; $\alpha \coloneqq 180\degr - \mathrm{scattering~angle}$) and exhibits a distinctive pattern. This curve, which is studied extensively in previous research \citep[e.g.,][and references therein]{1986MNRAS.218...75G}, can be divided into different regions. Near the opposition ($\alpha=0$), $\polr < 0$, where it reaches the minimum $\polr(\alpha=\amin) = \Pmin <0$ at $\amin \lesssim 10\degr$. Then, $\polr$ becomes positive after the inversion angle ($\ainv \sim 15 \thru 30\degr$) and reaches the maximum $\polr(\alpha=\amax) = \Pmax$ at $\amax \sim 90\thru150\degr$. Consequently, the region with $\alpha<\ainv$, where $\polr < 0$, is referred to as the negative polarization branch (NPB), while the $\alpha>\ainv$ region is known as the positive polarization branch (PPB).

PPC has been extensively studied since the pioneering works of \cite{1929PhDT.........9L} and \cite{1955PhDT........37D}. Several physical explanations for the NPB have been proposed, most notably the coherent backscattering mechanism \citep[][and references therein]{1994EM&P...65..201S,2002aste.book..123M,2015aste.book..151B, Hapke2012}.
Empirically, it has been shown that the NPB is influenced by factors such as particle size (\citealt{1990MNRAS.245...46G}), albedo contrast and/or mixing \citep{1987SvAL...13..182S, 2022A&A...665A..49S}, multiple reflections of absorbing particles \citep{1956AnAp...19...83D,1971LPSC....2.2285D}, and specifically, double reflections along the side-way direction (\citealt{1975ApOpt..14.1395W,1980Icar...44..780W,1981ApOpt..20.2493W}, \citealt{1984MNRAS.210...89G}; Figure 13.17 in \citealt{Hapke2012}), as well as large-scale surface roughness and/or compaction (\citealt{1970GeCAS...1.2127G}; \citealt{2002Icar..159..396S}; Appendix of \citealt{2022A&A...665A..49S}). Many theoretical improvements have proposed that the real part of the refractive index and interparticle distance at given wavelengths, as well as the particle size distribution, are important parameters \citep{Muinonen2002NATO,2009AJ....138.1557M}. Among these factors, the effect of particle size can be used to indirectly measure particle size and complement other size estimation methods, such as thermal modeling.

Moreover, after qualitative discovery by \cite{Umow1905}, the tendency toward stronger polarization for lower albedo is called the Umov law or the Umov effect \citep[e.g.,][\S 13.3.2]{Hapke2012}. However, the albedo-$\Pmax$ relation varies based on several physical factors, notably particle size \citep{1986MNRAS.218...75G,1986Icar...67...37D,1993JGR....98.3413D}. Another important relation between the albedo and polarimetric slope ($h \coloneqq d\polr/d\alpha(\alpha=\ainv) $; first noted in \citealt{1967AnWiD..27..109W}) seems to follow a tight relation regardless of the object \citep[to name few]{1973Icar...19..230V, 1973LPSC....4.3167B, 1986MNRAS.218...75G, 2015MNRAS.451.3473C, 2018SoSyR..52...98L}. There is another correlation between albedo and $\Pmin$ \citep[e.g.,][]{2015MNRAS.451.3473C,2018SoSyR..52...98L}. For simplicity, we refer to both the $h$-albedo and $\Pmin$-albedo relations as Umov law(s) in this paper. Specifically, the former is colloquially called the ``slope-albedo law,'' a name that has likely been used since \cite{1973LPSC....4.3167B}. These relations have been widely used to estimate the albedo of asteroids from polarimetric observations \citep{2018SoSyR..52...98L,2022MNRAS.509.4128I,2022MNRAS.516L..53G,2023MNRAS.525L..17G}.

In this study, we focus on investigating the NPB through an empirical approach. We emphasize that the experimental data are scattered across diverse literature sources, with variations in the definitions of albedo, polarimetric parameters, sample preparation methods, and experimental conditions. While compiling the sources, we recognized typos and errors in the figures and tables sourced from cross-referenced literature. Additionally, several crucial trends identified in this work received limited coverage in the original literature. As a result, our initial efforts were concentrated on investigating, collecting, and consolidating the published experimental datasets and extracting a trend that is practically useful for grain size estimation.

We compiled existing experimental data from published sources and excluded results from theoretical studies. We limited our focus to (i) the NPB of linear polarization (ii) under unpolarized incident light by (iii) a thick powdery surface of (iv) solid non-icy, non-metallic surface samples with (v) reasonable accuracy. For accuracy, as a rule of thumb, we used $\Delta\Pmin \lesssim 0.3\pp$, $\Delta \ainv \lesssim 2\degr$, etc., as criteria, depending on the shape of the PPCs. Icy surfaces are an important subject of study, especially for outer solar system objects. However, it is challenging to keep them frozen without sintering during the measurements, and the PPC depends on the regolith property below the ice (\citealt{1994MNRAS.271..343D}; \citealt{2023Icar..39615503S}), which complicates the interpretation. Despite their transparency, glass powder samples are discussed multiple times due to the wealth of available experimental results.

We also did not extensively explore the polarization opposition effect (POE), which is believed to arise from the coherent backscattering phenomenon (\citealt{1989SoSyR..23..111S}; \citealt{1990PhDT.......329M}; Section 13.4 of \citealt{Hapke2012}). The POE was reported as early as \cite{1929PhDT.........9L} for a MgO coating, leading to the identification of a distinct sharp negative polarization peak at very small $\alpha$ ($\sim 1\degr$). This effect significantly contributes to the asymmetry of the NPB. While it is possible to explore the POE by introducing skewness-related parameters, we opted not to do so due to the limited availability of samples with skewness information. This limitation stems from the fact that the original publications either could not reliably measure the full PPC at small phase angles due to experimental constraints or reported only $\Pmin$ without presenting the complete PPC. We only qualitatively described POE when necessary.

We first describe the data we collected (Sect. \ref{s:exp data}). Then, using these data, we revisit the Umov laws ($h$--albedo and $\Pmin$--albedo; Sect. \ref{s:umov}), including lunar and asteroid observations, to obtain the first ideas of the correlations. Next, we discuss the effect of fine particles a few times larger than the observation wavelength (Sect. \ref{s:wd trend}). Afterward, we disentangle the possible effects of albedo, compression, and the refractive index (Sect. \ref{s:other effects}). Equipped with this, we try to quantify the effect of fine particles, now down to the subwavelength scale (Sect. \ref{s:wd-nd transition}). The essence of the methodology is summarized in Sect. \ref{s:summ_trends} (Fig. \ref{fig:trends-schema}). Afterward, we justified our strategy developed throughout this work by estimating the grain sizes of selected celestial objects (Sect. \ref{s:application}). Finally, we briefly discuss how this work complements an independent thermal modeling approach (Sect. \ref{s:thermal model}) and provide a future perspective (Sect. \ref{s:conclusion}).

We also present our observational data for (1) Ceres and (4) Vesta in multiple figures, as well as optical data from published resources. However, detailed discussions of these airless objects are provided in the companion paper (hereafter, Bach et al. 2024; {\PaperII}). Thus, we do not discuss them in this work, although they are shown in the figures.

\section{Experimental data}\label{s:exp data}
There has been an enormous amount of effort put into experimental studies to understand the PPC of celestial objects. \cite{2015psps.book...62L} provides a thorough review of multiple instruments and results, especially for PROGRA$^2$[-Surf] \citep{2000P&SS...48..493W, 2009JQSRT.110.1755H}.
Table \ref{tab: exp data} summarizes the samples and original publications that are most relied on in this work. The collected datasets are briefly described in Sect. \ref{ss:exp data general}. Due to the importance of the alumina experiment (\citealt{1990MNRAS.245...46G}; {\Alumina} hereafter), this experiment is described in detail in Sect. \ref{ss:exp data alumina}, and the refractive index of alumina is discussed in Sect. \ref{ss:n of alumina}
for use in later parts of the study

\begin{table}[!tb]
\caption{Sources of experimental data.}
\label{tab: exp data}
\centering
\begin{tabular} {ll}
\hline \hline
Name\tablefootmark{a} & Sources \\
\hline
Lunar fine & \textbf{T0}, DG75  \\
Lunar dusty rock & \textbf{T0}, DG75\\
Lunar dust-free rock & \textbf{T0}, DG75\\
Meteo. Dust-free & \textbf{T0}, Z77a, S84\\
Meteo. Powder & \textbf{T0}, Z77a, BZ80, S84, LB89\\
Terr. Dust-free & \textbf{T0}, Z77a\\
Terr. Rock & L29\\
Lyot's powders & L29\\
Terr. ($>50 \um$) & \textbf{T0}, L29, Z77a, LB89\\
Terr. ($<25\thru50 \um$) & \textbf{T0} \\
Terr. ($<25 \um$) & Z77a\\
Terr. ($<10 \um$) & S04, S06\\
Volcanic Ash & L29, Z77a, S04 \\
Candle soot & Z77a \\
Rubber soot & S02 \\
Silicates + carbon & Z77b \\
Alumina[+ carbon] & {\Alumina} \\
Transparent glass powder & S92, S94, S02 \\
Soot + MgO & S92 \\
Soot[+ MgO] & B05\tablefootmark{b} \\
Mineral mix & LB89\tablefootmark{c}, M21, S22, S23 \\
Silica mix & S23 \\
\hline
\end{tabular}
\tablefoot{
\tablefoottext{a}{``Terr.'': Terrestrial, ``Meteo.'': Meteoritic.}
\tablefoottext{b}{One soot and one soot + MgO mixture (2 samples only).}
\tablefoottext{c}{Only a single olivine sample from the source.}
}
\tablebib{L29: \cite{1929PhDT.........9L};
DG75: \cite{1975LPSC....6.2749D};
Z77a: \cite{1977LPSC....8.1091Z};
Z77b: \cite{1977LPSC....8.1111Z};
BZ80: \cite{1980Icar...43..172L};
S84: \cite{1984SvAL...10..331S};
GD86: \cite{1986MNRAS.218...75G};
LB89: \cite{1989Icar...78..395L};
GG90: \cite{1990MNRAS.245...46G};
S92: \cite{1992Icar...95..283S};
S94: \cite{1994EM&P...65..201S};
S02: \cite{2002Icar..159..396S};
S04: \cite{2004JQSRT..88..267S};
S06: \cite{2006JQSRT.100..340S};
B05: \cite{2005Icar..178..213B};
M21: \cite{2021ApJS..256...17M};
S22: \cite{2022A&A...665A..49S};
S23: \cite{2023Icar..39515492S}.}
\end{table}

\subsection{Description of compiled data} \label{ss:exp data general}
Different publications used different definitions of albedo, making direct data comparison challenging. The two most widely used methods are denoted MgO5 (albedo measured at $\alpha=5\degr$ with respect to a thick MgO-smoked plate) and H2 ($\alpha=2\degr$ with respect to compressed Halon). For the former, MgO5-N indicates that the publication explicitly mentioned the use of normal incidence when illuminating the MgO plate. Similarly, the definition of slope $h$ has at least two widely used definitions: linear regression to $\alpha \in [\ainv,\, \ainv+10\degr]$ and fitting a second-order polynomial (``parabola'') for data at $\alpha \in [\ainv - 5\degr,\, \ainv + 10\degr]$. We call the former the 0-10-1 slope and the latter the 5-10-2 slope. We note that the exact definition of $h$ changes its value significantly for some PPCs.

We primarily used \cite{1986MNRAS.218...75G} as the main reference. These authors included measurements made only at the ``orange'' filter ($\lambda=5800\Angstrom$, half-width $\sim 500\Angstrom$). The tables provide tabulated values based on the Meudon Observatory log book. We call their tabulated data \textbf{T0}. In \textbf{T0}, the PPCs were measured at the specular geometry, and the albedos are the MgO5-N albedos. The uncertainties are $\Delta \Pmin = 0.05 \pp$, $\Delta \Pmax = 0.1 \pp$, $\Delta A = 0.01$, and $\Delta h = 0.005 \,\%/\degr$, but the uncertainties for each sample were not provided. The slope $h$ values in \textbf{T0} are in the 0-10-1 scheme.

Given that the table (\textbf{T0}) and figures in \cite{1986MNRAS.218...75G} do not match for some samples \citep{2019JKAS...52...71B}, we relied on the values provided in \textbf{T0}, not the figures. We carefully cross-checked the numeric values in \textbf{T0} against those from the original references. If the differences were negligible (roughly the size of the markers in the figures shown in this work), we prioritized the values presented in \textbf{T0}. If the differences were nonnegligible, we treated them as separate samples.

More lunar or terrestrial sample data (including volcanic ash) from \cite{1975LPSC....6.2749D} and \cite{1977LPSC....8.1091Z} are included. Terrestrial samples from the latter are explicitly described as $< 25\um$, so we labeled them ``Terr. ($< 25\um$)'', even though \textbf{T0} used the notation ``smaller than $50\um$''.

Many terrestrial samples were obtained from \cite{1929PhDT.........9L}, especially from Table 15. Since they do not describe whether they removed dust from the rocks, we carefully cross-matched samples in this and \textbf{T0} samples, separated them from the dust-free sample, and labeled them simply ``Terr. rocks''. These results are likely comparable to those of the  ``lunar dusty rock'' samples. Transparent crystal samples (e.g., NaCl) and three mineral powder samples (magnetite, pyrite, and galena) are collectively labeled ``Lyot's powder''. The study also included powdered lava samples which were reported to exhibit size $D=100\thru200\um$. The ``Volcanic ash samples included a wide range of grain sizes.

The PPCs and Table 15 of \cite{1929PhDT.........9L} are carefully compared with the values reported in \textbf{T0}. We found and fixed a few typos and mismatches. For mismatches, we prioritized the PPC: The value that better matches the PPC was adopted. When both of the tabulated values were clearly different from the PPC (e.g., $\Pmax$ is reported from hand-drawn curves, not from the real data points, and they differ greatly), we fixed it by visual inspection of the actual data points. Although the definition of albedo reported is not clear\footnote{
  On Page 101 of the English version, the albedo is defined by ``the surfaces of the two [the sample and white screen of magnesia] being normally illuminated and examined along the directions close to the normal.''
}, we assumed it was a MgO5 scheme under $\lambda=0.580\um$ \citep[see][]{1986MNRAS.218...75G}.

All the crushed meteorite samples from \cite{1977LPSC....8.1091Z} are included. The authors generated ``crushed but not ground or sieved'' meteorite samples to simulate cratering on asteroids, and the resulting grain sizes were reported to be $\lesssim 500 \um$. We marked these samples as ``Meteo. Powder'' for simplicity because the exact grain sizes are unavailable. The volcanic ash data (with only qualitative sample descriptions) are also adopted. The slope $h$ values are measured under the 5-10-2 scheme.

\cite{1980Icar...43..172L} contains PPCs and/or verbal explanations of eucrites (Moama, Bereba, ``Bereba B'' mixture, and Padvarninkai mixture) but lacks quantitative values for polarimetric parameters. Thus, based on the data presented in the figures or text, we extracted $h$, $\amin$, $\ainv$, and $\Pmin$ information, but albedo information was unavailable in most cases.

\cite{1989Icar...78..395L} contains 13 meteoritic, five terrestrial, and five metal samples, with grain sizes of $\le 50\um$. The authors used the MgO5-N albedo scheme for the freshly deposited MgO plate. 

\cite{1984SvAL...10..331S} reported PPCs of pulverized (size $D<100\um$) and cleaved faces of Allende and Kainsaz. We recovered polarimetric parameters by visual inspection. The albedos are measured at $\alpha=3\degr$ and $\lambda=0.56\um$. The authors described that the cleavage face ``do not exhibit either the slightest trace of any dust coat or any evidence of complex microrelief.'' Thus, we included cleavage samples in the ``Dust-free'' category.

\cite{1977LPSC....8.1111Z} provided valuable data for artificial silicates, described ``nearly amorphous hydrated magnesian silicate, $\mathrm{MgO\cdot SiO_2} x \mathrm{H_2O}$,'' with added carbon and/or black paint. We have included all their data ($\Pmin$, 5-10-2 slope $h$, MgO5-N albedo, $\ainv$). We also added $\amin$ by visual inspection. Carbon particles are reported to have a size of $0.01\um$, but no information is given for silicate. Considering the description in \cite{1979aste.book..170D}, which reports that, ``The powders made of pulverized silicates ... are strongly depleted in micron-size grains when compared to the lunar soil,'' we expect the particle sizes for the silicate powder to be much larger than those of the carbon particles (especially compared to $10\um$ silica in \citealt{2022A&A...665A..49S}).

The two ``candle soot'' datasets of the MgO5-N albedo (1.1\%) are adopted from \cite{1977LPSC....8.1091Z}. These are described as ``deposited on tray'' and ``loosely piled''. Another soot (rubber soot) dataset with a H2 albedo of 2.6\% is obtained by fitting to the PPC in \cite{2002Icar..159..396S}; this sample is prepared by incinerating black rubber and particle sizes $D\ll 1\um$.

\cite{1992Icar...95..283S} shows the transparent glass powder and soot-MgO mixtures. Since we could not access the quantitative albedo and grain size information, only $\Pmin$-$\ainv$ information was reconstructed.
Similarly, we extracted as much information as possible from \cite{2002Icar..159..396S} and included other glass powders in this work, combining them into ``transparent glass powder''. The albedo information is available only as ``moderate albedo, 40\thru60\%''. Figure 23 of \cite{1994EM&P...65..201S} shows black, green, and clear glass powders in the $\log_{10}(D)$-$\Pmin$ and $\log_{10}(D)$--$\ainv$ spaces. Among them, we recovered the ($\Pmin$, $\ainv$) pair only for the clear glass samples because the number of samples in the two panels differs for the other glass samples.

\cite{1987SvAL...13..182S} reported the PPCs of glass, soot, MgO, and $\mathrm{Fe_2O_3}$ fine powder (size $D<1\um$) and their mixtures. They argued that the NPB develops when the contrast in albedo between the endmembers increases. The PPCs of $\mathrm{Fe_2O_3}$ and soot are different from those of \cite{2002Icar..159..396S} and \cite{2005Icar..178..213B}, respectively, which is likely because of the different samples and preparation steps. We did not include the samples presented in this work but  qualitatively describe them when necessary.

\cite{2004JQSRT..88..267S} provided some valuable experimental data for fine samples (with sizes ranging from $D=2\thru14 \um$). We calculated $\Pmin$, $\ainv$, $\amin$, and $h$ by visual inspection, and we noted the possibility of greater uncertainty. The figures for feldspar and loess are identical in the publication, likely by mistake, although albedos seem to be correctly written in the figures. Based on the slope-albedo law, we concluded that the figure for the loess mistakenly overlie the feldspar one. The blue-wavelength PPC corresponding to feldspar could be restored from Figure 14 in \cite{2006JQSRT.100..340S}. Although \cite{2007JQSRT.106..487S} shows the full results for feldspar, the NPBs are printed at low resolution, and possibly because of this, some $\ainv$ values do not coincide with the previous values for the same sample. Their albedos are in the H2 scheme \citep{2004JQSRT..88..267S}.

\cite{1992Icar...95..283S} and \cite{2005Icar..178..213B} report soot and MgO mixture results. The former shows multiple samples with $\Pmin$ and $\ainv$, while the latter provides two PPCs of pure soot (albedo 3.4\%) and a soot+MgO(5\% wt) mixture (albedo 4.3\%). The definition of albedos is not given, but it is likely the H2 scheme.

\cite{2022A&A...665A..49S} conducted a significant study on the mixing effect of minerals. In their work, the illuminated region on the sample was approximately $15\,\mathrm{mm}$ in diameter under $\lambda=0.53\um$, and the ``albedo'' adopted in this work was hemispherical-directional reflectance. All the powder samples had grain sizes $D<20 \um$, although silica formed larger aggregates. Their SEM images show that the grain sizes are comparable to the wavelength scales, especially for the forsterite and magnetite samples. Given the important information provided by of their samples and high-quality data, we have collected their PPCs \footnote{
Spadaccia (2023; priv. comm.), \url{https://zenodo.org/record/7550710}
}
and obtained the corresponding NPB parameters, including the 5-10-2 scheme $h$ values. Each of the fitted curves was visually inspected, and our $\ainv$ values matched well with the reported values within the error bars; therefore, we adopted $\ainv$ from the publication. We carefully compared the available PPC and the published values and corrected two typos ($\ainv$).

\cite{2023Icar..39515492S} reported a series of hyperfine particle experiments in which Mg-rich (forsterite) olivine was used $D=0.6\um$ and iron sulfide (FeS) was used $D=0.2 \um$ (see Table 1 and Supplementary Figure 3). Polarimetry was conducted under the normal emission geometry at $\lambda=0.53\um$, and the albedo is the solar spectrum-averaged ``equivalent'' albedo estimated from $\lambda=0.56\um$ reflectance\footnote{
Poch (2023, priv. comm.), \url{https://zenodo.org/record/7649167}, and Beck (2023, priv. comm.).
} (\citealt{2021Icar..35414066B}). After collecting their PPCs, we calculated the NPB parameters, especially the 5-10-2 $h$ and $\ainv$ values, as above.

\subsection{Description of the alumina experiment (\Alumina)} \label{ss:exp data alumina}
In this work, and its underlying logic, the alumina sample data (\Alumina) play a pivotal role. We collected $\Pmin$, $\ainv$, and MgO-5 albedo values, if available, from the publication. Their measurements were conducted using four filters attached to a Minipol mounted on a B. H. Zellner's goniometer, denoted as B (blue), G (green), R (red), and I (infrared), corresponding to peak wavelengths of $\lambda = 0.440,\, 0.535,\, 0.685,\, 0.790\um$, respectively. These filters were applied in conjunction with various particle sizes ($D=40,\, 12,\, 3,\, 1,\, 0.3,\, 0.05 \um$), resulting in a total of nine distinct $D/\lambda$ measurements spanning the range from $0.093$ to $75$. The samples were on a blackened stainless steel, $4\,\mathrm{mm}$ depth, ``pressed flat with a spatula'', observed at the specular geometry, rotated by motor to ``average out the effects of any specular reflections from grain facets'', so that $D/\lambda$ is the only parameter changing over various measurements.

Notably, among these measurements, those for $D/\lambda=75,\, 22,\, 5.6,\, 1.9,\, 0.56,$ and $0.093$ were conducted with the same G filter. This unique set of measurements provides an opportunity to isolate the effects of the refractive index and grain size. Additionally, for $D/\lambda=75,\, 0.56,$ and $0.093$, carbon particles ($D=0.01\um$) were added to alter the albedo of the samples, with all the measurements conducted within the G filter.

\begin{figure}[tb!]
\centering
\includegraphics[width=\linewidth]{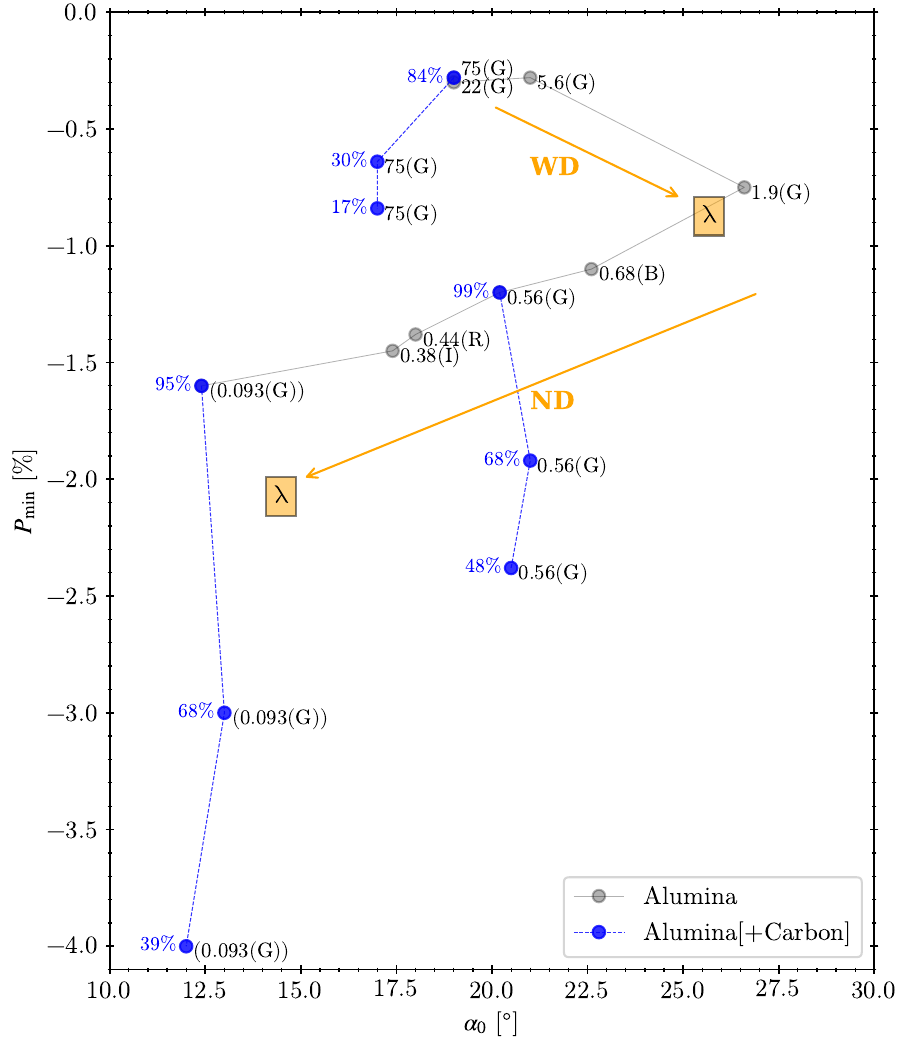}
\caption{Laboratory samples in $\Pmin$-$\ainv$ (depth-width of NPB) space. The numbers to the right of the alumina indicate $D/\lambda$ and the filter, while the blue numbers to the left (with unit $\%$) are the albedo values. $D/\lambda = 0.093$ data points are enclosed in parentheses because their $\Pmin$ values are extrapolated values. The WD and ND trends, with increasing $\lambda$ (or decrease $D/\lambda$), are schematically indicated.}
\label{fig:pmin-a0-lab-simple}
\end{figure}

Their results are summarized in the $\Pmin$-$\ainv$ space in Fig. \ref{fig:pmin-a0-lab-simple}. The authors found the following trend in the NPB as the wavelength ($\lambda$) increased:
\begin{enumerate}
\item Unchanged when $D/\lambda \gg 5 $ (samples of $D/\lambda=22, \, 75$).
\item Wider and deeper (WD, hereafter) when $D/\lambda \sim 5 \rightarrow 2$.
\item Narrower and deeper (ND, hereafter) when $D/\lambda \lesssim 2$.
\end{enumerate}
In their study, $\Pmin$ values for samples with $D/\lambda = 0.093$ are estimated through extrapolation; thus, we consider these values to be uncertain. Furthermore, their PPCs exhibit a notably sharp NPB at small $\alpha$ when $D/\lambda \lesssim 1$, which is indicative of a strong POE, even for these flattened samples. While it would be ideal to treat the broad NPB and the POE separately, for the purposes of this work, we simply use $\Pmin$ and $\ainv$ in our analysis. These findings suggest that the ND trend may be closely associated with the POE and coherent scattering characteristics (\citealt{1989SoSyR..23..111S}; \citealt{1990PhDT.......329M}; Section 13.4 of \citealt{Hapke2012}).

Because $h$ values were not provided in \Alumina, we digitized the PPC values for $\alpha < 40 \degr$. Because of the limited number of data points, we could not calculate $h$ under the 0-10-1 or 5-10-2 schemes. Rather, we fit a linear line to the six data points nearest to $\ainv$. This approach was further validated by ensuring that the $\ainv$ derived from this linear fit remained consistent with the original values reported in the publication.

\subsection{Refractive index of alumina}\label{ss:n of alumina}
The refractive index $n = n(\lambda)$ is known to affect the NPB \citep{2002ocd..conf..261M}. Consequently, for future discussion, we have compiled information on $n(\lambda)$ for alumina (aluminum oxide, $\mathrm{Al_2O_3}$, also called [synthetic] sapphire in the literature) reported in various sources.

\subsubsection{Wavelength dependence}
One of the first reported values under ordinary rays was tabulated in \cite{1962JOSA...52.1377M}, and a dispersion relation was proposed to be established at $24\mathrm{\degr C}$ for $\lambda=0.26\thru5.5 \um$ ($\lambda$ in $\mathrm{\mu m}$):
\begin{equation*}
    n_o^2 - 1 =
      \frac{1.023798\lambda^2}{\lambda^2 - 0.06144821^2}
      + \frac{1.058264\lambda^2}{\lambda^2 - 0.1106997^2}
      + \frac{5.280792\lambda^2}{\lambda^2 - 17.92656^2}
     ~.
\end{equation*}
The selected measurements are
\begin{equation*}
\begin{aligned}
  n_o(\lambda= 0.43 \um) &= 1.78199 ~, &&
  n_o(\lambda= 0.53 \um) = 1.77191 ~, \\
  n_o(\lambda= 0.69 \um) &= 1.76362 ~, &&
  n_o(\lambda= 0.79 \um) = 1.76046 ~.
\end{aligned}
\end{equation*}

At an annual meeting of the Optical Society of America, an updated version was presented \citep{1972_Malitson}. At $20\mathrm{\degr C}$, for $\lambda = 0.20\thru 5.0\um$ (recited from TABLE II of \cite{1997_Tropf_Alumina} and Section 1.3.4 of \citealt{2002_Weber}):
\begin{equation*}
\begin{aligned}
    n_o^2 - 1 &=
      \frac{1.43134936\lambda^2}{\lambda^2 - 0.0726631^2}
      + \frac{0.65054713\lambda^2}{\lambda^2 - 0.1193242^2}
      + \frac{5.3414021\lambda^2}{\lambda^2 - 18.028251^2} ~, \\
    n_e^2 - 1 &=
          \frac{1.5039759\lambda^2}{\lambda^2 - 0.0740288^2}
          + \frac{0.55069141\lambda^2}{\lambda^2 - 0.1216529^2}
          + \frac{6.59273791\lambda^2}{\lambda^2 - 20.072248^2} ~,
\end{aligned}
\end{equation*}
for the ordinary and extraordinary rays, respectively. At selected wavelengths, the calculated values are
\begin{equation*}
\begin{aligned}
  (n_o, n_e) (\lambda=0.43\um) &= (1.78191, 1.77355)~, \\
  (n_o, n_e) (\lambda=0.44\um) &= (1.78059, 1.77226)~, \\
  (n_o, n_e) (\lambda=0.79\um) &= (1.76040, 1.75247)~.
\end{aligned}
\end{equation*}

A later report by \cite{1985umo..rept.....Q} tabulated the $n(\lambda)$ value for polarizations parallel ($n_\parallel$) and perpendicular ($n_\perp$) to the crystal $c$-axis. The values at the two selected wavelengths are:
\begin{equation*}
\begin{aligned}
  n_\parallel(\lambda=0.44\um) &= 1.730 (\pm 0.010) + i 0.020 (\pm 0.001) ~,\\
  n_\perp(\lambda=0.44\um)     &= 1.754 (\pm 0.011) + i 0.021 (\pm 0.001) ~,\\
  n_\parallel(\lambda=0.79\um) &= 1.727 (\pm 0.020) + i 0.020 (\pm 0.010) ~,\\
  n_\perp(\lambda=0.79\um)     &= 1.750 (\pm 0.011) + i 0.020 (\pm 0.001) ~.
\end{aligned}
\end{equation*}
We note that these values deviate largely from those of other works.

\begin{figure}[tb!]
\centering
\includegraphics[width=1\linewidth]{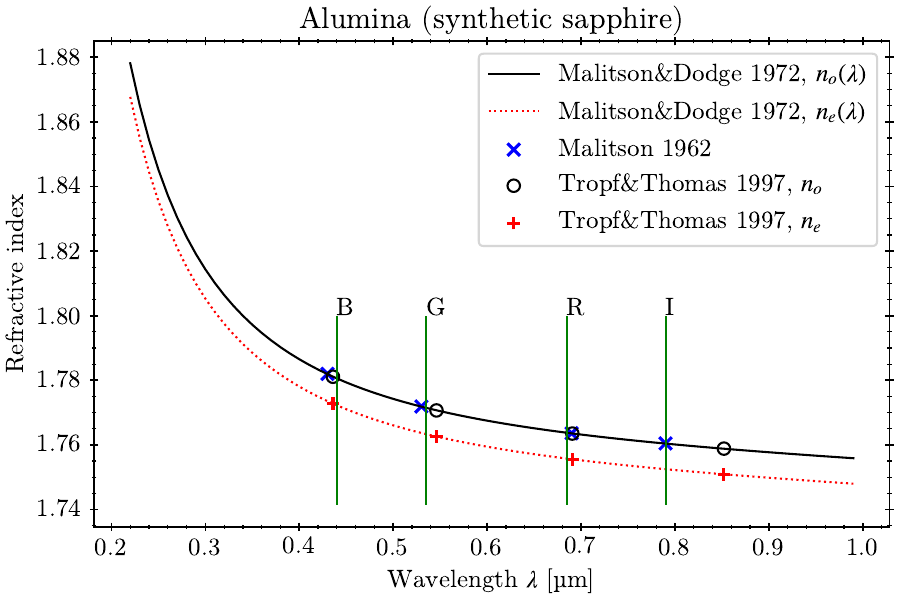}
\caption{Refractive indices of alumina ($\mathrm{Al_2O_3}$) from various sources (see text for details). Available measurements closest to the central wavelengths of the filter (B, G, R, and I used in \Alumina) are indicated with markers.}
\label{fig:alumina-ref-ind}
\end{figure}

\cite{1997_Tropf_Alumina} reported $n(\lambda)$ for ordinary ($n_o$) and extraordinary ($n_e$) rays. Selection from their work:
\begin{equation*}
\begin{aligned}
  (n_o, n_e)(\lambda = 0.43583\um) &= (1.78110, 1.77275)~, \\
  (n_o, n_e)(\lambda = 0.54607\um) &= (1.77067, 1.76254)~, \\
  (n_o, n_e)(\lambda = 0.69072\um) &= (1.76351, 1.75549)~, \\
  (n_o, n_e)(\lambda = 0.85212\um) &= (1.75885, 1.75090)~.
\end{aligned}
\end{equation*}
Combining these (except inconsistent values from \citealt{1985umo..rept.....Q}), the refractive index of alumina is decreased by $\sim 0.02$ from B ($0.44\um$) to I ($0.79\um$) filter used in \Alumina, as shown in Fig. \ref{fig:alumina-ref-ind}.

\subsubsection{Temperature dependence}
In regard to the temperature dependence of the refractive index, $n_T (T) \coloneqq \frac{dn}{dT}$ and \cite{1975JSSCh..12..213V} reported that, under the ``room temperature (RT)'',
\begin{equation*}
    (n_{T o}, n_{T e})(\mathrm{RT}, 0.589\um) = (13.6, 14.7)\times 10^{-6}\,\mathrm{[K^{-1}]}
\end{equation*}
for ordinary and extraordinary rays, respectively.

\cite{1986_Dodge} (recited from \citealt{2002_Weber} Section 1.3.5) reported that, at $\lambda=0.4579\um$:
\begin{equation*}
\begin{aligned}
  (n_{T o}, n_{T e})(-180\mathrm{\degr C}, 0.4579\um) &= (1.8, 1.9)\times 10^{-6} \,\mathrm{[K^{-1}]} ~,\\
  (n_{T o}, n_{T e})(20\mathrm{\degr C}, 0.4579\um) &= (11.7, 12.8)\times 10^{-6} \,\mathrm{[K^{-1}]} ~,\\
  (n_{T o}, n_{T e})(200\mathrm{\degr C}, 0.4579\um) &= (15.4, 16.9)\times 10^{-6} \,\mathrm{[K^{-1}]} ~.
\end{aligned}
\end{equation*}

\cite{1986JOSAA...3..610T} reported a fitting formula for the ordinary ray under $T=24\thru1060\mathrm{\degr C}$ (to the 0.02\% with 99\% confidence level):
\begin{equation*}
\begin{aligned}
  n_{T o}(T, 0.633\um) &= 1.76565 + 1.258\times 10^{-5} T + 4.06\times10^{-9} T^2\\
  n_{T o}(T, 0.799\um) &= 1.75991 + 1.229\times 10^{-5} T + 3.10\times10^{-9} T^2\\
\end{aligned}
\end{equation*}
where the temperature $T$ is $\mathrm{\degr C}$.

Therefore, for a main-belt asteroid with a diurnal temperature variation on the order of $\Delta T = 10^2 \mathrm{K}$, the expected change ($n_T$) is on the order of $10^{-3}$ throughout the wavelength range of interest. For the related alumina experiment, the wavelength-dependent change is already an order of magnitude larger than that related to temperature, and $\Delta T$ in the experiments must be very small, such that $n_T$ can be neglected in our analyses.

\section{A first look: The Umov laws (albedo-$h$-$\Pmin$)} \label{s:umov}

\subsection{The correlations}
\begin{figure*}[!tb]
\centering
\includegraphics[width=0.7\linewidth]{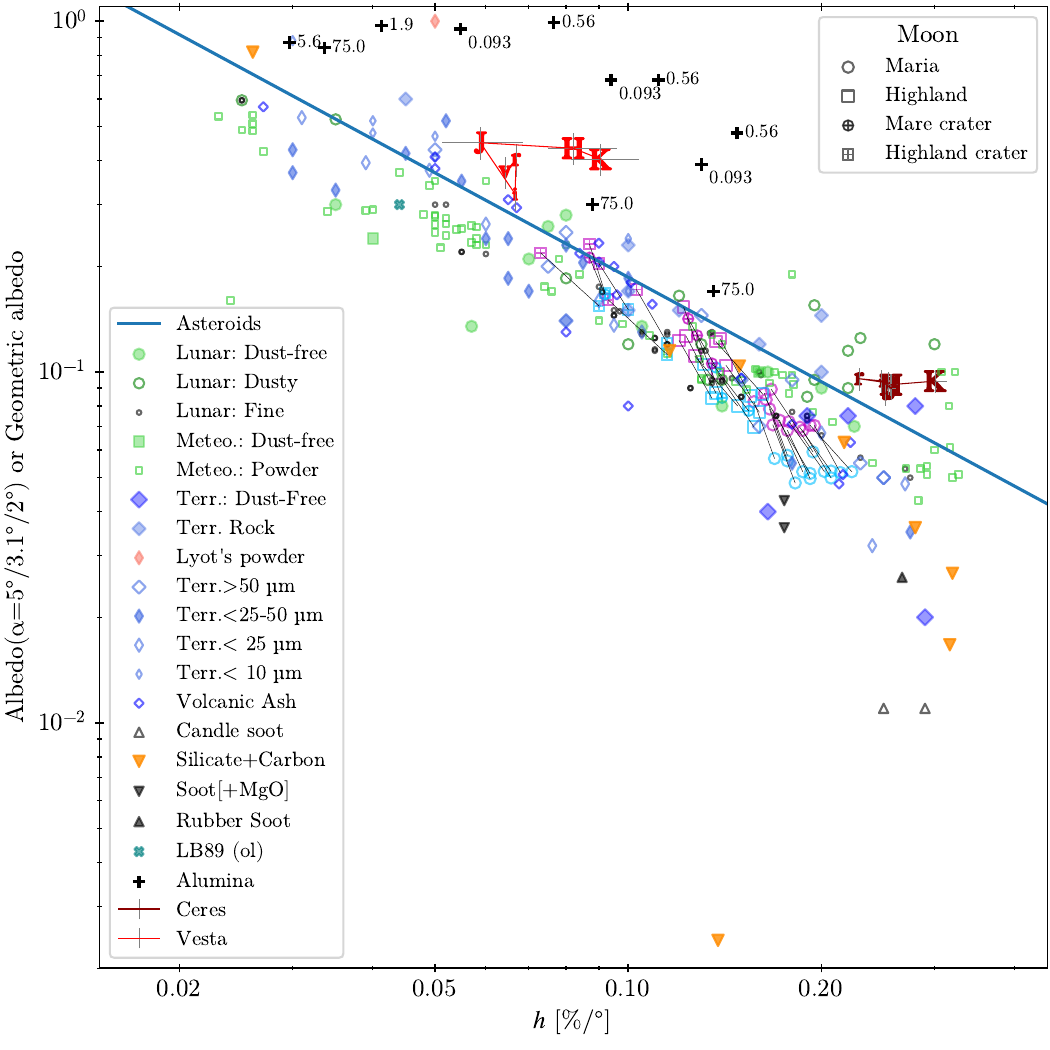}
\caption{Slope ($h$)-albedo relation. See Sect. \ref{s:exp data} and Table \ref{tab: exp data} for the details of the laboratory data and their original sources. The numbers next to the alumina samples are the $D/\lambda$ values. The telescopic lunar observation data from \cite{1992Icar...95..283S} are shown for blue ($0.42\um$) and red ($0.65 \um$) wavelengths using cyan and magenta colors, respectively, and the same regions are connected by a thin line. The lunar albedo is defined at $ \alpha=3.1\degr$, so it should be smaller than the geometric albedo. The blue solid line is the regression line for asteroids \citep{2018SoSyR..52...98L}. The Ceres and Vesta results are shown as red and dark red, respectively, and are connected in the order of increasing wavelength: v=$0.45\thru0.60\um$, r=$0.60\thru0.75\um$, i=$0.75\thru1.0\um$, J-, H-, and K-bands (\PaperII).}
\label{fig:h-albedo}
\end{figure*}

Among the Umov laws, the slope-albedo relation is well known for its tightness. In Fig. \ref{fig:h-albedo}, we plot the compiled samples in the $h$-albedo space. The overturn of the linear relation, as evidenced by candle soot and silicate-carbon mixtures, starts to appear at albedo $\lesssim 5\,\%$. \cite{1977LPSC....8.1111Z} explained this with the extremely reduced amount of double scattering by dark inclusions, based on \cite{1975ApOpt..14.1395W}. This trend has also been found for asteroids \citep{2015MNRAS.451.3473C}. Figure 4 in \cite{1992Icar...95..283S} also shows the overturning at low albedo for water-color samples. For high-albedo objects, the $h$-albedo correlation appears to become less distinct.

\begin{figure*}[!tb]
\centering
\includegraphics[width=0.7\linewidth]{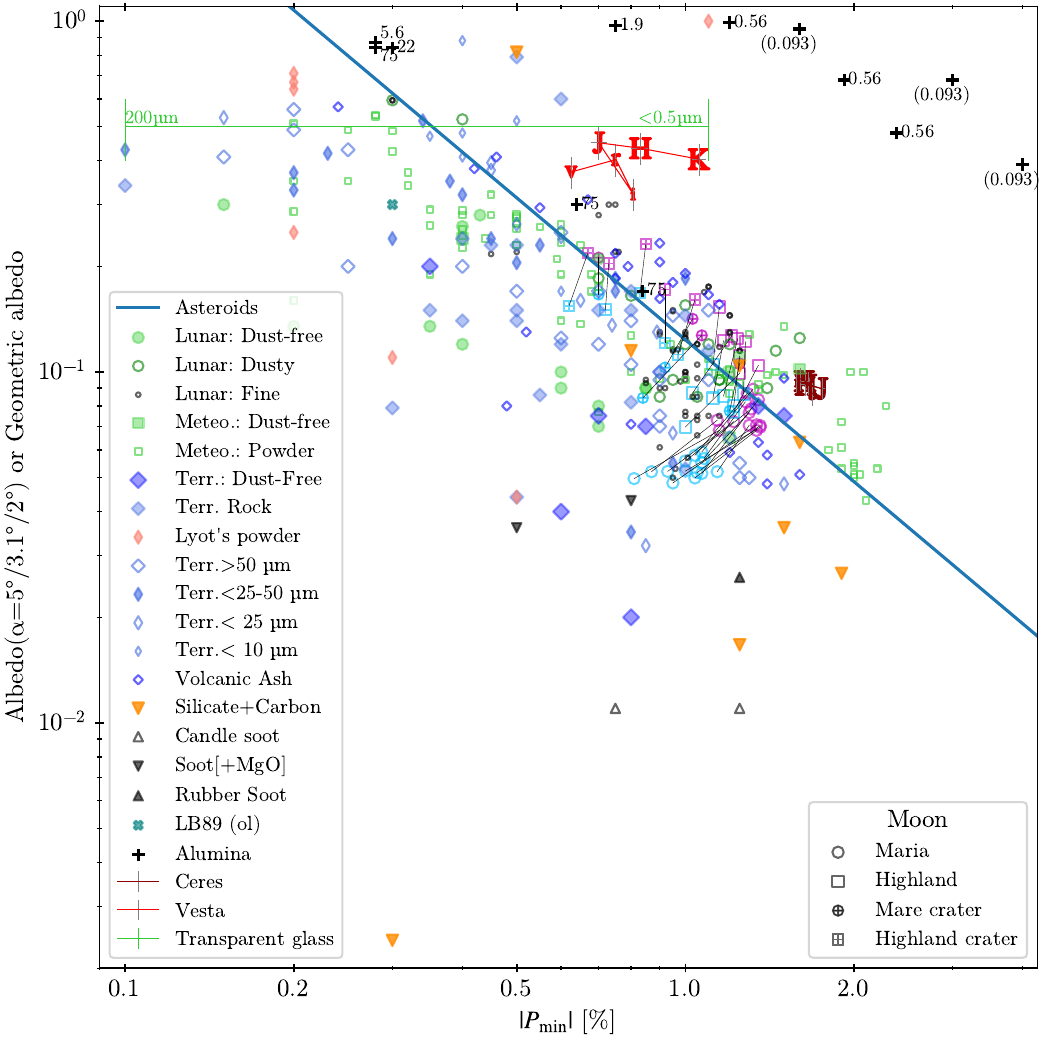}
\caption{$|\Pmin|$--albedo relation. The legends are identical to those in Fig. \ref{fig:h-albedo}. The transparent glass powder samples are shown as a range, with the grain size ($D$) at the ends indicated because the albedo for each sample was unavailable. For lunar observations, $\Pmin = \polr(\alpha=10.5\degr)$ is assumed \citep{1992Icar...95..283S}. For alumina samples, $D/\lambda=0.093$ is enclosed in parentheses because their $\Pmin$ values are extrapolated values (\Alumina).}
\label{fig:pmin-albedo}
\end{figure*}

The $\Pmin$-albedo relation, shown in Fig. \ref{fig:pmin-albedo}, is known to be more scattered than the slope-albedo law (e.g., \citealt{1979aste.book..170D} for experiments and \citealt{2015MNRAS.451.3473C} and \citealt{2018SoSyR..52...98L} for asteroid observations). One logical hypothesis is that there is a strong secondary parameter that controls this $\Pmin$-albedo relation, such as particle size, compaction, and roughness. Like in the slope-albedo relation, $\Pmin$-albedo relation also shows an overturn or departure from the general trend in the lowest and highest albedo regions.

The general asteroid trend in the figures (blue solid lines from \citealt{2018SoSyR..52...98L}) shows the geometric albedo. Thus, asteroid ``albedo'' values must be systematically larger than those of laboratory samples due to the opposition surge. Additionally, the albedo of lunar observation \citep{1992Icar...95..283S} is measured at $\alpha=3.1\degr$, while assuming that $\amin = 10.5\degr$, that is, $\Pmin = \polr(\alpha=10.5\degr)$. By comparing the tables and figures for the lunar observation, we found some mismatches; therefore, we consider the table values; because mare craters and highland craters are oppositely labeled, we correct for this.

The two-band lunar observations in Fig. \ref{fig:h-albedo} show a small systematic shift in the trend for the two wavelengths. In red, $h$ is larger (steeper) than expected at the same albedo in blue. The lunar data in Fig. \ref{fig:pmin-albedo} exhibit a behavior that diverges from the typical trend observed for asteroids. The trend even appears to run in the opposite direction in the case of maria.

\begin{figure*}[tb!]
\centering
\includegraphics[width=1\linewidth]{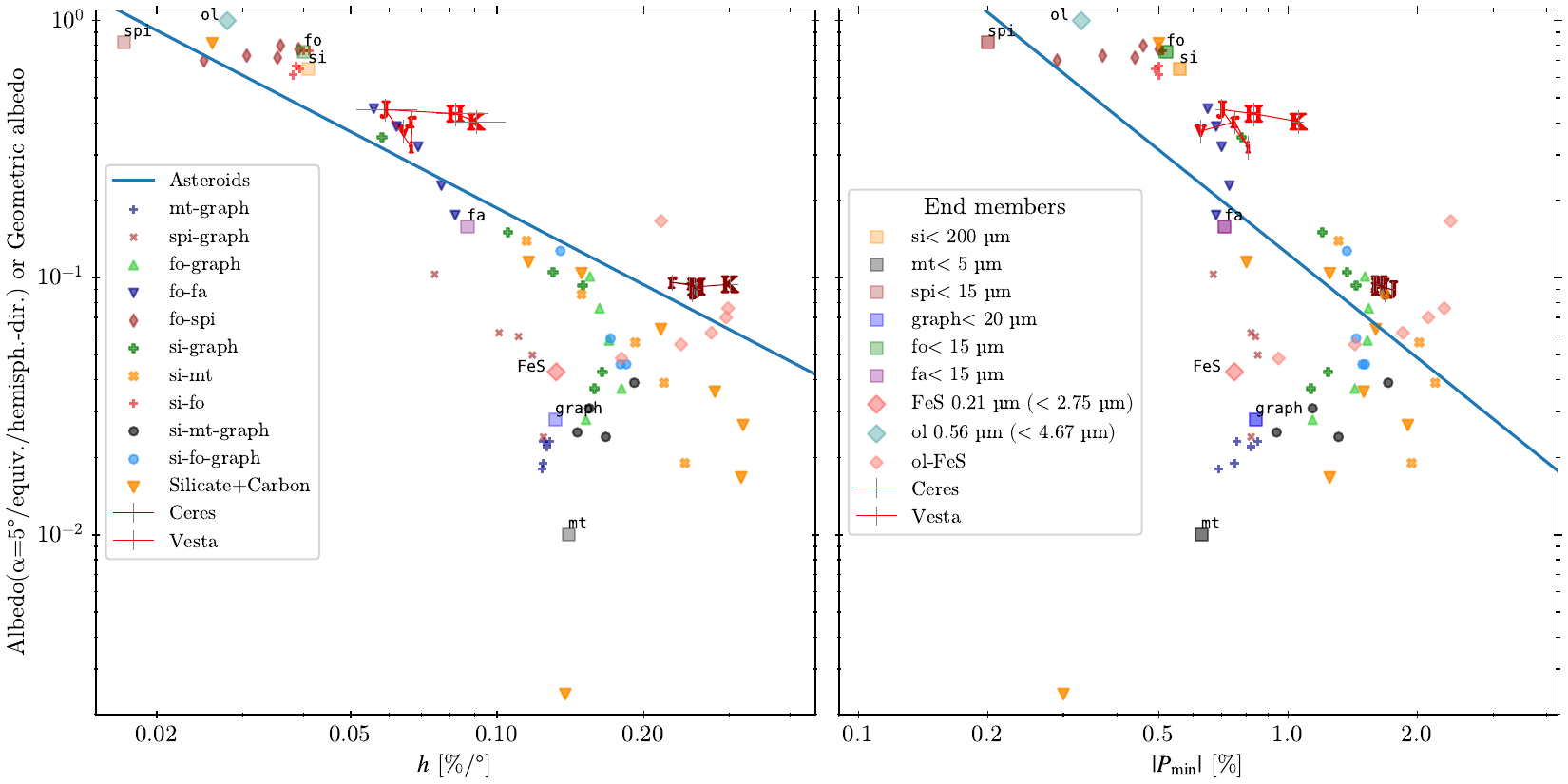}
\caption{Umov laws for mineral and silica mixture samples that use different albedo definitions. The square and diamond endmembers are from \cite{2022A&A...665A..49S} and \cite{2023Icar..39515492S}, respectively (see Sect. \ref{s:exp data} for details). Silicate-carbon mixture samples and Ceres and Vesta are shown for comparison (see Fig. \ref{fig:h-albedo}). The abbreviations are as follows: si: silica, mt: magnetite, spi: spinel, graph: graphite, fo: forsterite, fa: fayalite, ol: olivine, and FeS: iron sulfide. Binary and tertiary mixtures are indicated by connecting endmembers with a hyphen (``-'') in the legend.}
\label{fig:umov-minerals}
\end{figure*}

Fig. \ref{fig:umov-minerals} shows identical relationships but for the mineral and silica mixture samples. Their albedos are defined differently from those in Figs. \ref{fig:h-albedo} and \ref{fig:pmin-albedo}, so they cannot be compared directly. Olivine samples and mixtures (ol, fo, fa) lie on a smooth curve, even though they are from two independent studies, strengthening the credibility of the experiments.

\subsection{Samples worth noting}
The distribution of alumina in $D/\lambda < 2$ is largely different from that in the other $D/\lambda>5$ samples in both parameter spaces. {\Alumina} noted that each group of samples with $D/\lambda = 75,\, 0.56$ and $ 0.093 $ qualitatively follow the Umov law in the $\Pmin$-albedo space: higher albedo samples have shallower $\Pmin$ (they did not describe slope $h$). The transparent glass powder samples in Fig. \ref{fig:pmin-albedo} also show a deepening trend as the grain size decreases to the scale of $\lambda$, similar to alumina. Additionally, although not plotted, $D<1\um$ glass powder samples of $67\%$ and $89\%$ H2 albedos \citep{1987SvAL...13..182S} are located near alumina samples of $D/\lambda=0.56$ or $1.9$ in both parameter spaces.

There are two outliers among high albedo terrestrial rock samples in Fig. \ref{fig:pmin-albedo}: The chalk\footnote{
  Rotationally averaged, 7 mm spherically cut sample, $79\%$ albedo, $\Pmin=-0.5\%$, $\amin=15\degr$, $\ainv=31\degr$, $\Pmax=5\%$.
}
and the white clay\footnote{
  $58\%$ albedo, $\Pmin=-0.6\%$, $\amin=3\degr$, $\ainv=10\degr$, and $\Pmax=21.7\%$ were obtained from Yonne, France. \cite{1986MNRAS.218...75G} reports slightly different values, so we adopt these original values based on the PPC.
}. The original report \citep{1929PhDT.........9L} mentioned that chalk has a similar PPB shape to that of precipitated chalk\footnote{
  Obtained from $\mathrm{CaCl_2}$ and $\mathrm{Na_2CO_3}$ solutions.
}. For the white clay, the report described it is ``somewhat remarkable'' to show such a deep NPB for its albedo, though no further description is given. The NPB shape of the white clay (sharp minimum at $\amin=3\degr$) is similar to that of small alumina samples (\Alumina), which hints at the possibility of the existence of fine particles.

Among ``Lyot's powder'' samples, the magnesia coating\footnote{
  MgO, 100\% albedo, $\Pmin=-1.1\%$, $\amin=0.7\degr$, $\ainv=23\degr$, $\Pmax=8.2\%$ \citep{1929PhDT.........9L}. Although not plotted, the 0-10-1 slope $h\sim0.05\%/\degr$ from Figure 2 of \cite{2002Icar..159..396S}.
} is worth attention. The original report noted that the sharp $\Pmin$ at a very small $\amin$ is attributed to the fine particle size of the sample. A micrograph of the MgO coating showed a similar PPC \citep[Figure 3 of][]{2002Icar..159..396S} and indeed confirmed the presence of a plethora of sub$\lambda$ particles in the sample. Notably, the MgO samples from both works are located near the alumina sample of $D/\lambda<1$.

As shown in Fig. \ref{fig:umov-minerals}, the spi-graph mixture has a smaller $h$ and shallower NPB than the other samples with similar albedos. However, ol-FeS mixtures reach higher $h$ and deeper NPB than other samples of similar albedo, which may be linked to their specific fine, sub$\lambda$ scale grain sizes. For very fine samples ($D < \lambda$), the $\ainv$ and $|\Pmin|$ of mixtures are greater than those of both endmembers \citep{1987SvAL...13..182S}. Interestingly, both $h$ and $\Pmin$ change greatly when albedo is fixed for the fo-spi mix, similar to glass powder (Fig. \ref{fig:pmin-albedo}). Generally, the mixtures continuously change between the endmembers. Other samples also roughly follow the typical trends expected by the Umov law.

\section{Grain size ($D/\lambda$) effect: The widening and deepening (WD) trend}\label{s:wd trend}
The size parameter $X \coloneqq \pi D/\lambda$ is a controlling parameter for scattering processes once the refractive index is known \citep{1983asls.book.....B}.
For an observer, the target object has a fixed $D$, so changing $\lambda$ leads to an adjustment of $X$; thus, the scattering process will be affected.
To easily compare with the experimental studies (especially \Alumina), we use $D/\lambda $ ($= X/\pi$), rather than $X$.

\begin{figure*}[!tb]
\centering
\includegraphics[width=0.8\linewidth]{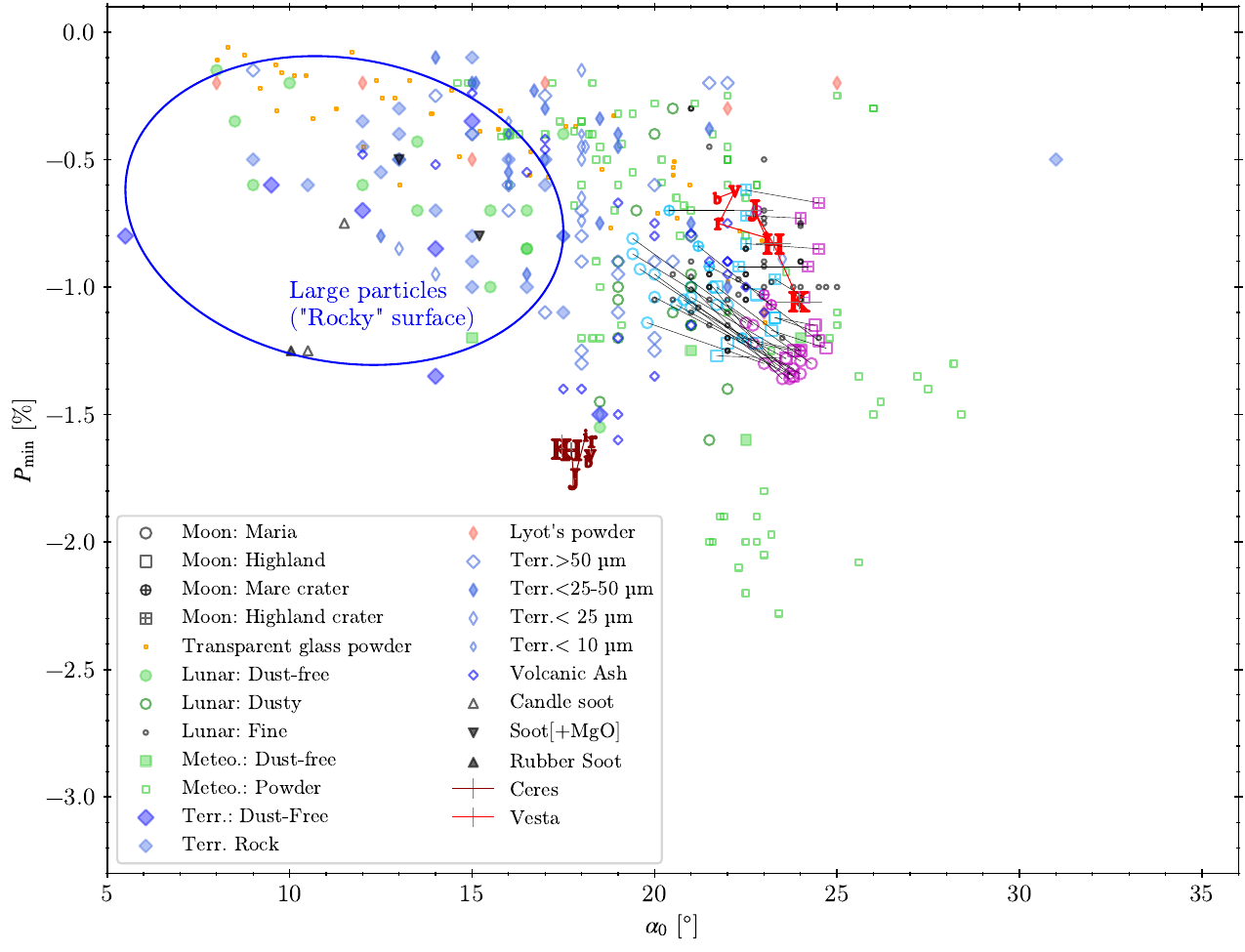}
\caption{$\Pmin$-$\ainv$ parameter space plot for samples and lunar observations. The legends are the same as those in Fig. \ref{fig:pmin-albedo}, and the typical location of rocky surfaces is indicated with a blue ellipse. Ceres and Vesta are shown for comparison (see Fig. \ref{fig:h-albedo}; b=$0.30\thru0.45\um$).}
\label{fig:pmin-a0-sample}
\end{figure*}

$\Pmin$ and $\ainv$ describe the shape of the NPB in the first order (in terms of depth and width, respectively). In Fig. \ref{fig:pmin-a0-sample}, we plotted the compiled samples and lunar observations in the parameter space. As proposed previously, particle size indeed changes NPB-related parameters: fine particle samples are well separated from other dust-free rock samples. Smaller particle samples generally show a wider and often deeper NPB \citep{1977RSPTA.285..397D}. \cite{1989aste.conf..594D} used this diagram to argue that asteroids are covered with coarser particles than on the Moon.

Below, we further discuss the evidence supporting the WD trend, including the widening and deepening trend with decreasing $D/\lambda$, based on previous experimental works and lunar observations. Then, we attempt to quantify at which $D/\lambda$ the WD trend starts to occur.

\subsection{Previous works}
Figure 12 in \cite{1961plsa.book..343D} shows the WD trend for iron filings as the particle size decreases. Figure 1 of \cite{1969A&A.....2..105D} shows widening (but no clear deepening) as the wavelength increases in the case of pulverized limonite. In a study by \cite{1967JGR....72.3105K}, changes in the NPB according to laboratory experiments were observed, with a deepening effect as $\lambda$ increased. However, further discussion was limited, which is potentially attributed to crude particle size segregation. Later, {\Alumina} conducted alumina powder experiments under varying $D/\lambda$ conditions (Sect. \ref{ss:exp data alumina}).

\cite{1992Icar...95..283S} briefly mentioned that they observed a qualitatively similar WD trend for transparent glass powders of size $D = 0.5\thru200 \um$ (Figs. \ref{fig:h-albedo} and \ref{fig:pmin-albedo}, and in Fig. \ref{fig:pmin-a0-sample}). In addition, this same WD trend (likely correlated with $D/\lambda$) is clearly observed for the green, clear, and black glass samples in Figure 23 of \cite{1994EM&P...65..201S}.

\cite{2002Icar..159..396S} exhibits alumina experiments in red ($\lambda=0.62 \um$) light, which show a qualitatively similar, deepening trend when $D$ is changed from $3.2 \um$ to $0.5$ ($ D/\lambda =5.2$ to $0.8$). Interestingly, the $D=0.1$ and $0.5\um$ samples showed nearly identical PPCs and $\Pmin$. However, a deepening trend was observed for both samples when $D/\lambda$ was changed by changing $\lambda$ from $0.45$ to $0.63\um$ (their Figure 16), which coincided with the deepening as $D/\lambda$ decreased. Because this deepening trend is also nearly identical for both samples, one possible hypothesis for explaining the cause of the trend is that both samples share a similar particle size frequency distribution, despite being labeled with different sizes.

\cite{2002P&SS...50..849N} reported the deepening trend for $D=1.2\um$ alumina samples from $\lambda=0.545$ to $0.633\um$ ($D/\lambda=2.2$ to $ 1.9$). Because of the limited $\alpha$ range, the width trend cannot be extracted from these studies. We note that they concluded that the wavelength dependence predicted by coherent backscattering theory was not observed.

\cite{2015Icar..250...83D} also showed a marginal WD trend when $\lambda$ is increased ($0.488$ to $0.647\um$) for the JSC200 and calcite samples, although all the trends were not conclusively distinguishable under the given error bars. Additionally, their size frequency distributions indicate the possibility of a plethora of sub$\lambda$ grains, which could complicate the interpretation of our findings. Under a microgravity environment, \cite{2023MNRAS.520.1963H} reported a similar result: NPB becomes increasingly wider and deeper (WD) when smaller ($<50\um$) particles are used.

Although this is an experiment with a metallic substance, Figure 13.15 of \cite{Hapke2012} depicts a comparable WD trend for iron spheres as their size diminishes from $D=20$ to $3\um$ ($D/\lambda \sim 5$). Similarly, although an icy (frost) surface is not the main focus of this work because of its transparent nature, we note that the WD trend becomes more visible as the ice frost size decreases \citep{1994MNRAS.271..343D,2018JGRE..123.2564P}

The qualitative observations from these studies suggest a strong connection between the NPB and the WD trend in the presence of fine particles. Therefore, we conclude that the WD trend is likely not only a characteristic of the alumina sample but also a trend that applies more generally.

\subsection{The $D/\lambda$ upper bound for the WD trend}\label{ss:wd quanti}
We showed the WD trend is observed across a broad spectrum of samples, not limited to alumina (Sect. \ref{ss:exp data alumina}; \Alumina). Now we review some additional publications that provide insights leading us toward a quantification of the upper boundary for the $D/\lambda$ range within which the WD trend is observed.

A study by \cite{2021ApJS..256...17M} investigated the polarimetric properties of forsterite powders, encompassing a size range from $0.7$ to $\sim 100\um$, alongside $1600 \um$ pebbles. The emergence of the WD trend is observed as the particle size decreases, commencing at $D/\lambda \sim 5\thru10$.

\cite{2022MNRAS.517.5463F} found that the NPB tends to become slightly stronger (the WD trend) for smaller particles in both olivine and spinel samples. The size frequency distribution of the samples (in their Figure 2) indicates that their ``small'' sample corresponds to particles $D \sim 2 \um$ ($D/\lambda \sim 4 $), while their ``medium'' sample encompasses the ``small'' particles along with additional particles $D \sim 25\um$. This mixture of particle sizes may contribute to the less distinct WD trend (i.e., small and medium samples both show similar NPB) in their study.

Figure 12 in \cite{2001JGR...10617375V} presents an analysis of various aerosol samples, including red clay, quartz, Pinatubo ash, loess, and Lokon ash. The figure shows the WD trend for these samples as the wavelength increases from $0.4416\um$ to $0.6328\um$. Notably, despite these samples having overall effective grain sizes of $D\lesssim 10 \um$, the strongest peaks in size-frequency distributions occur at $D \sim 1\thru2 \um$ (as shown in their Figure 1). This observation supported the existence of grains with values of $D/\lambda$ that became conducive for the WD trend to occur.

\cite{2018ApJS..235...19E} demonstrated that the NPB significantly diminishes when small (radius $< 1\um$) particles are removed from lunar simulants. We estimated that the number of removed particles was $0.5 \um < D \ll 10\um$ from their size frequency distributions and microscopy images. Conversely, the addition of $1 < D/\lambda \ll 20$ particles induces a WD trend.

Combining all these results with those of the previous section, while also considering the properties of {\Alumina} and context provided in our previous discussions, we propose that the WD trend is observed when $D/\lambda \lesssim 5\thru10$. This trend can be a diagnostic indicator of the upper bound of the grain size on the scattering surface upon observing it with increasing $\lambda$. However, we acknowledge that this threshold value is rather uncertain and subjective. Further research can provide a clearer understanding of the exact dependencies in various scenarios. For example, the location where the ND trend starts to appear, the WD--ND transition point, is discussed in Sect. \ref{s:wd-nd transition} after the other effects (especially that of albedo) are discussed.

\section{Other Effects on the negative branch ($\Pmin$ and $\ainv$)}\label{s:other effects}
Having explored the influence of $D/\lambda$ on the NPB, we focus on other factors. Specifically, we discuss the effects of albedo, compression, and the refractive index.

\subsection{Albedo}\label{ss:effect alb}
\begin{figure*}[tb!]
\centering
\includegraphics[width=\linewidth]{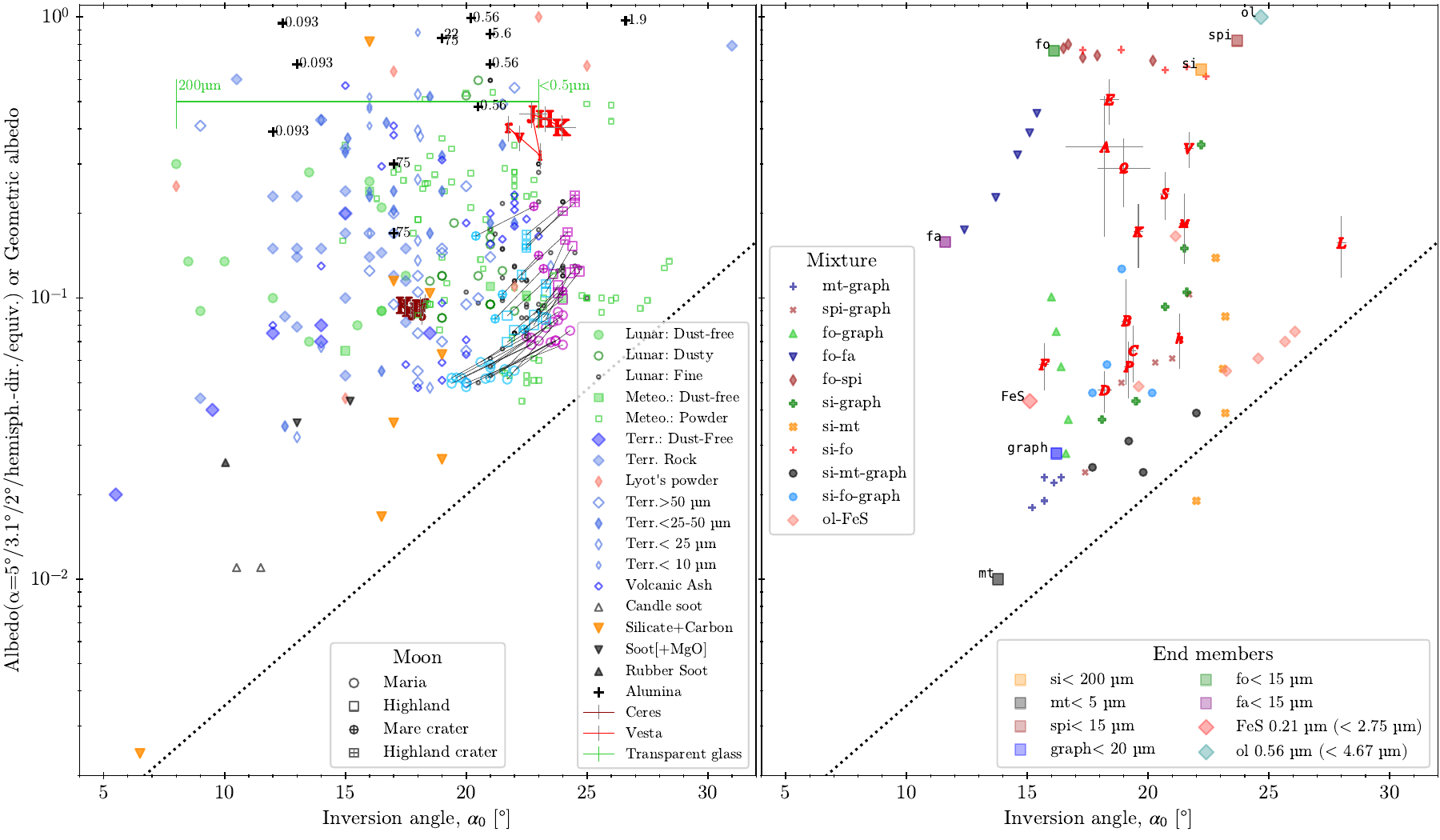}
\caption{Albedo-$\ainv$ relation. The legends on the left and right panels are identical to those in Figs. \ref{fig:h-albedo} and \ref{fig:umov-minerals}, respectively. The red letters on the right indicate the taxonomic types of asteroids of the letter, except ``h'' stands for the Ch-type \citep{2017Icar..284...30B}. The ``boundary'' (Eq. \ref{eq:alb-a0 boundary}) is indicated by black dotted lines. Ceres and Vesta are shown for comparison (see Fig. \ref{fig:h-albedo}).}
\label{fig:a0-albedo}
\end{figure*}

Fig. \ref{fig:a0-albedo} shows yet another parameter space, the albedo-$\ainv$ space, encompassing a diverse array of samples, including lunar observations \citep{1992Icar...95..283S} and asteroid data \citep{2017Icar..284...30B}. It is worth noting that while the specific numerical values may be arbitrary and the albedo definitions differ across studies, a discernible pattern emerges. We identify the following inequalities (the black dotted line in the figure):
\begin{equation}\label{eq:alb-a0 boundary}
    A \gtrsim 10^{0.075\ainv - 3.2}
    \quad\mathrm{or}\quad
    \alpha_\mathrm{0} \lesssim 47.7 + 13.3 \log_{10}A ~,
\end{equation}
for albedo $A$ in natural units and $\ainv$ in degrees.
Figure 4 in \cite{1992Icar...95..283S} also shows a similar ``boundary'' for water-color samples.

As already demonstrated in Fig. \ref{fig:pmin-a0-lab-simple}, $\ainv$ of alumina remains relatively constant with respect to albedo (for $\gtrsim 20\%$) but changes only if $D/\lambda$ is varied. The same is true for transparent glass powder samples (albedo $40\thru60\%$).
In the right panel of Fig. \ref{fig:a0-albedo}, it is shown that the mixture samples merely follow nearly vertical or L-shaped curves connecting their endmembers. The ``Silicate+Carbon'', fo-graph and si-mt mixtures show nearly fixed $\ainv$ over a wide range of albedos. Moreover, the fo-spi and si-fo mixtures exhibited an $\ainv$ change, although the albedo was nearly fixed. All these findings emphasize that the albedo is not the only factor tuning $\ainv$. Neither is the albedo contrast (e.g., Section 4.6 of \citealt{2022A&A...665A..49S}). From all the mixture samples, including those containing tertiary mixtures, we hypothesize that the minimum/maximum $\ainv$ is determined roughly by the endmember with the smallest/largest $\ainv$, except possibly for sub$\lambda$ scale endmembers \citep[e.g., the ol-FeS samples and][]{1987SvAL...13..182S}.
Further parameter spaces related to albedo and $\amin$ are briefly discussed in Appendix \ref{s:amin space}.

\begin{figure*}[bt!]
\centering
\includegraphics[width=0.8\linewidth]{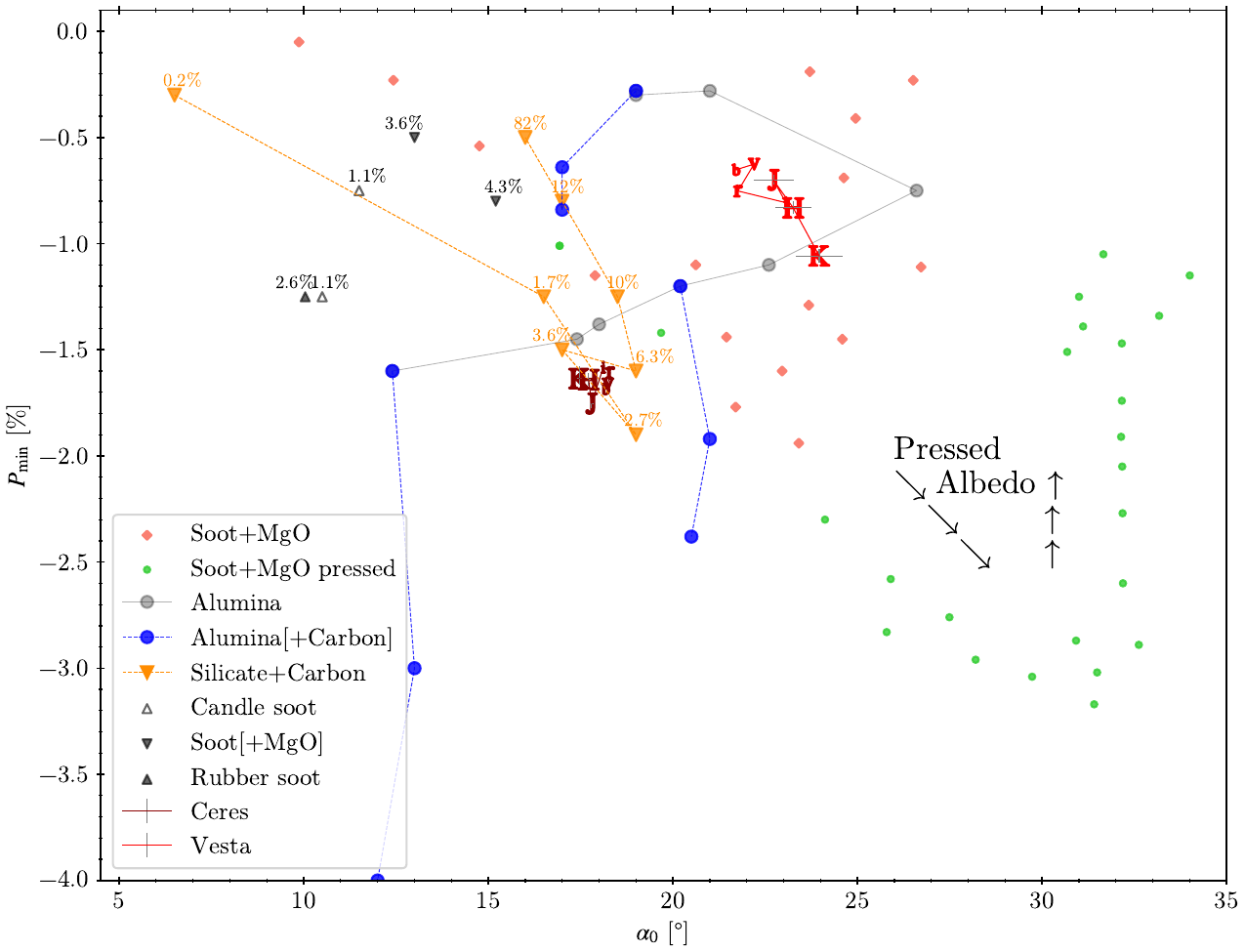}
\caption{Laboratory samples in $\Pmin$--$\ainv$ (depth-width of NPB) space. Alumina samples are shown in Fig. \ref{fig:pmin-a0-lab-simple}. The numbers with units $\%$ are albedo values. ``Silicate+Carbon'' samples are connected by lines in the order of albedo. The arrows for ``Pressed Albedo'' indicate the increasing albedo for the pressed MgO and soot mixture sample given in the original publication. Ceres and Vesta are shown for comparison (see Fig. \ref{fig:h-albedo}; b=$0.30\thru0.45\um$).}
\label{fig:pmin-a0-lab}
\end{figure*}

\begin{figure*}[tb!]
\centering
\includegraphics[width=0.8\linewidth]{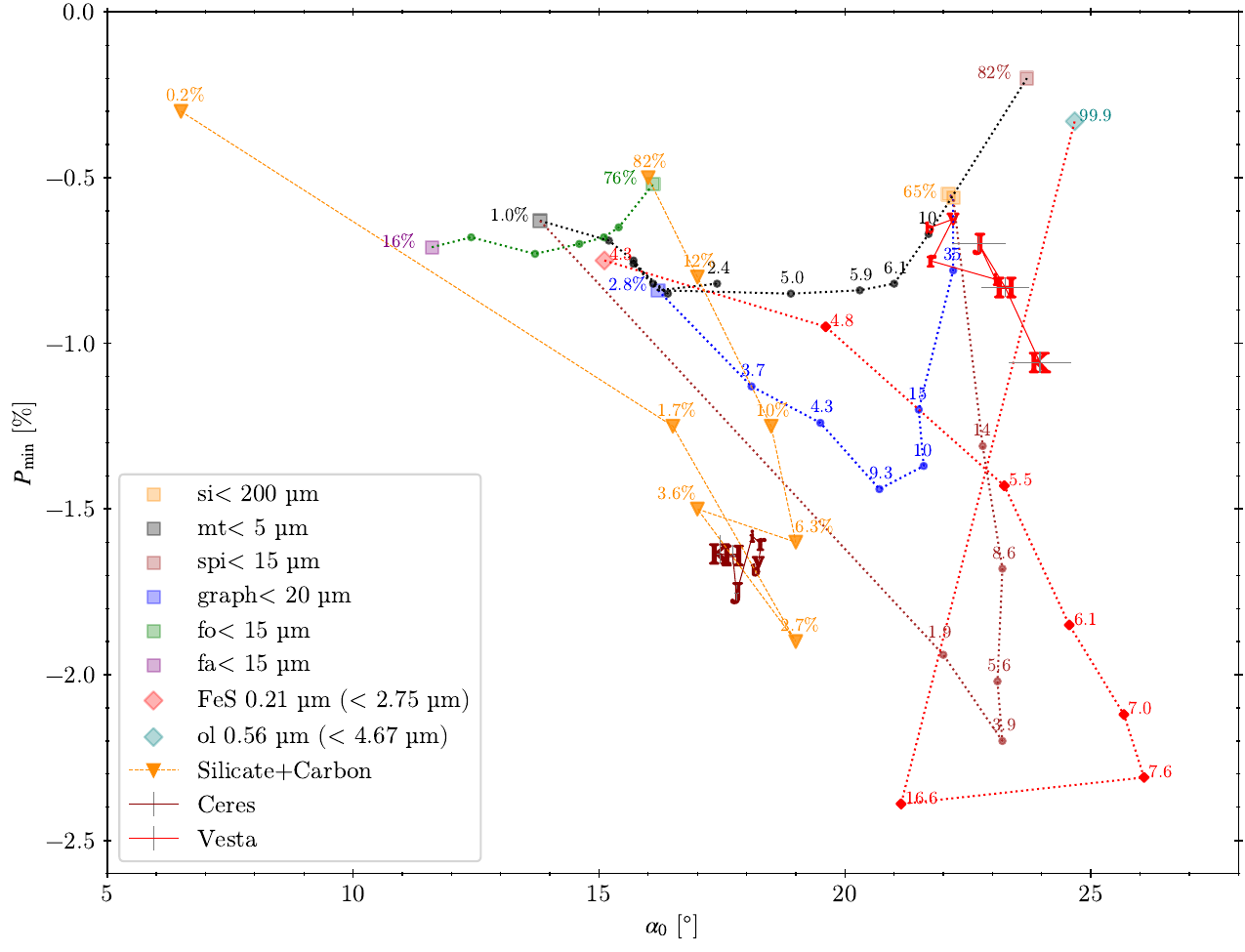}
\caption{Similar to Fig. \ref{fig:pmin-a0-lab} but different samples to emphasize the dependence of the NPB on albedo. Note that the axes ranges are slightly different. The endmembers are denoted by a filled square \citep{2022A&A...665A..49S} or filled diamond \citep{2023Icar..39515492S}. The dotted lines connect the mixtures in the order of albedo. The numbers indicate the reported ``albedo'' values. Ceres and Vesta are shown for comparison (see Fig. \ref{fig:h-albedo}).}
\label{fig:pmin-a0-spa22}
\end{figure*}

{\Alumina} described the effect of decreasing albedo on deepening the NPB, citing Figure 4 of \cite{1984MNRAS.210...89G} that it is theoretically expected. Figs. \ref{fig:pmin-a0-lab} and \ref{fig:pmin-a0-spa22} are drawn specifically to emphasize the effect of albedo. From the figures, as well as Fig. \ref{fig:a0-albedo}, one finds the following general trends for albedo:
\begin{enumerate}
\item For lower albedo ($\lesssim 2\thru10\%$), increasing albedo strengthens the NPB (WD-like).
\item With increasing albedo ($\gtrsim 5\thru10\%$), increasing albedo weakens the NPB (shallower $\Pmin$), which is the classical expectation of the Umov law, while $\ainv$ does not change much.
\end{enumerate}
We denote the first trend ``WD-like'' because we use the ``WD trend'' to indicate that $D/\lambda$ is the primary cause of the observation. Therefore, the albedo trend for low-albedo objects can be combined with the WD trend, complicating the analyses (Sect. \ref{ss:alb vs size}).

A qualitatively similar trend was reported in Figure 22 of \cite{1994EM&P...65..201S} for ``metallic or semimetallic rough pulverized'' samples and in Figure 4 of \cite{1992Icar...95..283S} for mixtures of water color ($\lambda=0.65\um$). According to \cite{2002Icar..159..396S}, the mixture of chalk and soot ($D < 1 \um$) at $\lambda=0.63 \um$ shows a deepening trend as adding soot decreases the albedo from 90\% to 8\%. This finding aligns with the albedo effect at higher albedos, as described above. The measured $\alpha$ range is limited in their work (up to $3.5\degr $), so the width ($\ainv$) trend cannot be mentioned.

Among the two ``Candle Soot'' samples, the shallower NPB is described as ``loosely piled'' sample. The sample (MgO5 albedo $1.1\%$) lies between the two ``Silicate+Carbon'' mixtures of albedos $0.2$ and $1.7 \%$, which indicates that albedo is the primary parameter contributing to these effects. Another ``Candle Soot'' with the same albedo (deeper NPB) deviates from the trend of ``Silicate+Carbon'' because it is a pure deposit, that is, not prepared by ``deliberately roughening\footnote{
\cite{1970GeCAS...1.2127G}, above Figure 9.
}'' procedure.

The single ``Rubber Soot'' data have an H2 albedo of $2.6\%$. The sample is likely a pure deposit soot sample, which may explain why it is located near the pure deposit candle soot. The two ``Soot[+MgO]'' samples are prepared by shaking the test tubes rather than roughening the surface\footnote{
The ``Rubber Soot'' and ``Soot[+MgO]'' samples have H2 albedos of $2.6\%$, $3.6\%$ and $4.3\%$ (5wt\% MgO). From the photometric phase curve of soot in Figure 20 of \cite{2002Icar..159..396S}, (albedo at $\alpha=5\degr$)/(albedo at $\alpha=2\degr) \sim 0.91 $, if the halon has a flat intensity phase curve from $\alpha=2\thru5\degr$. Applying this approach to all three samples, the MgO5-equivalent albedos are $2.4 \% $, $3.3 \%$, and $3.9\%$, respectively.
}. However, the two ``Soot[+MgO]'' samples still show a WD-like trend for increasing albedo, which is the trend for dark objects summarized above.

\subsection{Albedo vs. $D/\lambda$ effects}\label{ss:alb vs size}
As mentioned, the effects of albedo and $D/\lambda$ may play roles simultaneously, and both contribute to widening and/or deepening, which complicates the analyses. We look into a few cases that highlight this complication.

\cite{2002Icar..159..396S} reports interesting trends from two water colors and a $\mathrm{Fe_2 O_3}$ sample (all with $D \sim 1 \um$ particles). As the wavelength increased from $\lambda=0.45$ to $0.63\um$, the albedos and the NPB trends were reported as
\begin{itemize}
\item Red water color: $A=7$\% and 68\%, shallowing.
\item Blue water color: $A=43$\% and 24\%, deepening.
\item $\mathrm{Fe_2 O_3}$ (reddish): $A=5$\% and 20\%, no change.
\end{itemize}
The red water color follows the albedo trend of high-albedo objects. This is the opposite of the $D/\lambda$ effect, which is expected to deepen. This might indicate that the albedo effect dominates the $D/\lambda$ effect, as can be expected from its large change in albedo. Conversely, for the blue water color, deepening is anticipated from both the albedo and $D/\lambda$ effects, thus posing no contradiction. In the case of the $\mathrm{Fe_2 O_3}$ sample, it is possible that the shallowing effect driven by albedo and the deepening effect due to $D/\lambda$ counterbalance each other.

Among multiple samples, \cite{2001JGR...10617375V} reported that feldspar grains exhibit no change in $\Pmin$ when $\lambda = 0.4416\um$ is changed to $0.6328\um$. Their sample had a $D = 2 \um$ peak, which corresponds to $D/\lambda = 4.5$ and $3.3$ for the two wavelengths. They described feldspar as having a ``light pink'' color, which coincides with the $D=2\um$ feldspar sample albedos reported in \cite{2004JQSRT..88..267S}:
67\% and 88\% in $\lambda=0.45$ and $0.63\um$, respectively. Thus, similar to the $\mathrm{Fe_2 O_3}$ sample discussed above, the albedo effect (shallowing) and the $D/\lambda$ effect (deepening) may cancel each other out.

The fo-fa mixture in Fig. \ref{fig:pmin-a0-spa22} exhibits a trend different from that of the general one, a strong widening pattern, contrary to the other typical behaviors observed in high-albedo objects. The microscopy images \citep{2022A&A...665A..49S} show that forsterite contains significantly finer structures down to the wavelength scale than does fayalite. This result indicates that the trend observed in the fo-fa mixture might be influenced by the grain size effect, wherein a reduction in $D/\lambda$ toward forsterite contributes to widening, whereas the depth trend appears to be canceled out.

\subsection{Compression}\label{ss:effect rough}
The compression process plays a dual role in influencing the behavior of the NPB, with its impact characterized by the simultaneous reduction in macroscopic roughness (if present) and a consequential decrease in interparticle distance, which reduces porosity.
For example, when compressing ``Soot+MgO'', we observe a WD-like trend (Figure \ref{fig:pmin-a0-lab}). A similar trend was observed $15 \um$ for the SiC powder \citep[Figure 13.14 of][]{Hapke2012}. In both cases, the authors do not explicitly describe a roughening process, so compression affects porosity in this case.

The results presented in {\Alumina} are based on alumina samples flattened by a spatula to control all the other parameters except $D/\lambda$ (Sect. \ref{ss:exp data alumina}). The difference in the PPCs of different works using alumina samples (\Alumina, \citealt{2018JGRE..123.2564P}, \citealt{2002Icar..159..396S}, \citealt{2002P&SS...50..849N,2018Icar..302..483N}, \citealt{2006JQSRT.101..394O}) could also be due to the different processes of compression. The photomicrograph in \cite{2006JQSRT.101..394O} hints that compression can indeed change the shapes of aggregates.

The effect of compression on the MgO coating is reported in Figures 2 and 4 of \cite{2002Icar..159..396S}. A sharp peak with $\amin \lesssim 1\degr$ is exhibited only for uncompressed samples. After mechanical compression, $\Pmin$ became shallower, and $\amin$ and $\ainv$ increased (their Figure 2). After drying-in-alcohol compression, $\Pmin$ remains the same, and $\amin$ increases (likely $\ainv$, as shown in their Figure 4, which seems to be identical to the sample shown in Figure 2 in \citealt{Shkuratov2002NATO}).
This difference may be related to the fine-scale structure that existed in the MgO coating ($D\ll 1 \um$), which was responsible for the POE being removed after compression. We note that the availability of data points near the POE can distort the results. This shows the complication of NPB analyses for superfine particles unless we fully sample the PPC at a small size $\alpha$. Figure 13 in \cite{2002Icar..159..396S} and Figure 2 in \cite{Shkuratov2002NATO} also demonstrated that $D=0.01\um$ silica shows a WD-like trend after compression, but one needs data at $\alpha < 1\degr$, which is not available in most publications.

Inducing the WD-like trend is not the only possible effect of compression. For example, in the Appendix of \cite{2022A&A...665A..49S}, compression is shown to have the following additional effects:
\begin{itemize}
\item Silica: WD-like
\item Magnetite: Narrowing and shallowing
\item Silica-magnetite 1:1 mixture: Width fixed and deepening
\end{itemize}

Experiments with black plasticine sprinkled with dark glass powders showed a loss of NPB when the glass powders were carefully pressed into plasticine one by one \citep[Figure 13 in][]{1984MNRAS.210...89G}. In this specific example, ``compression'' resulted in the elimination of multiple scattering by dark particles. Similarly, flattening the lunar fine (Apollo 10084-6) resulted in a weakening of its NPB \citep[Figure 10 in][]{1970GeCAS...1.2127G}. Moreover, in the study of terrestrial samples, \cite{1971A&A....10..450D} noted that ``when the surface is flattened, the negative branch becomes less pronounced\footnote{
They also mentioned: ``Measures reported here are made with a surface roughened to such a degree that the $\Pmax$ is lowest.''
}''. \cite{1969A&A.....2..105D} described how compressing a limonite-goethite powder mixture (Mars simulant) resulted in shallower NPB, but there was no description about $\ainv$. All these findings are exactly the opposite of the WD-like behavior.

In the experiments involving lunar fine and terrestrial samples discussed here, compression involves not only a substantial decrease in porosity but also a notable reduction in surface roughness. This occurs because they compare ``deliberately roughened'' samples with corresponding pressed samples. Interestingly, \cite{1977LPSC....8.1091Z} reported that compaction did not ``have any noticeable effects'' on any of the meteoritic samples, except for the ``highly translucent Norton County aubrite.''

\subsection{Refractive index}\label{ss:ref ind}
Comparing Jovian satellites and lunar fines, {\Alumina} suggested that increasing the real part of the refractive index ($n$) may result in a WD-like trend. Earlier, \cite{1929PhDT.........9L} described grain shape as likely to be a more important factor than the refractive index by comparing powdered calcite and precipitated chalk, crushed glass and spherical glass powders with sodium chloride. In \cite{2022MNRAS.509.4128I}, a deepening trend due to hydration was found, which may be related to the refractive index.

The distributions of alumina in Figs. \ref{fig:h-albedo} and \ref{fig:pmin-albedo} may bear some resemblance to the theoretical expectation for metals (high imaginary refractive index; Figure 2 of \citealt{1980Icar...44..780W}), and one may conclude that the imaginary part of the refractive index is the controlling parameter. However, alumina is a nonmetallic substance. Furthermore, multiple samples with $D/\lambda=0.093\thru75$ were measured under the same filter (G), thus maintaining a constant refractive index yet still exhibiting deviations from the general trend. Thus, a change in the imaginary refractive index is unlikely to be the main cause of the difference between these samples and other samples.

Based on the findings of \cite{2009AJ....138.1557M}, \cite{2018MNRAS.479.3498D} argued that the change in $n$ of spinel (1.81 and 1.78 at $\lambda=0.42\um $ and $0.76\um$, respectively) may be affecting the narrowing trend of $\ainv(\lambda)$ of the Barbarian asteroids. Can this be the cause of the WD and/or ND trend in Fig. \ref{fig:pmin-a0-lab-simple}?

Alumina is anticipated to show $\Delta n \approx -0.02 $ (Sect. \ref{ss:n of alumina}) from the B to I band ($\Delta \lambda = 0.35\um$), while $\Delta \ainv \approx -5\degr$ in Fig. \ref{fig:pmin-a0-lab-simple}. \cite{2009AJ....138.1557M} showed that, according to the theory of \cite{Muinonen2002NATO}, the $\ainv$ value changes most significantly by $n$, but only a slight change is expected from the power-law particle size distribution. We first reproduced their results using the same formalism with a fixed power-law distribution with power $-2.5$, $n=1.77$, and minimum and maximum sizes $D\pm0.02\um$. We found that for $D=0.3\um$ under the G filter, $d\ainv/dn \approx +1\degr/0.01$. Thus, we expect $\Delta \ainv = -2\degr$ under the given $\Delta \lambda$. Second, if the case of spinel \citep[Figure 12 of][which gives $d\ainv/dn \sim +2\degr/0.01$]{2018MNRAS.479.3498D} is blindly applied, we expect $\Delta \ainv \sim -4\degr$. These two results appear to explain the observed ND trend to a certain degree. Moreover, in the case of spinel, $\Delta n \lesssim -0.02 $ from the V- to J-/H-bands \citep{2008PSSBR.245.2800H,2018MNRAS.479.3498D}, so it is expected that $\Delta \ainv \lesssim -4 \degr$. This interpretation coincides with the later observations of spinel-bearing asteroids in the J- and H-bands \citep{2023PSJ.....4...93M}.

However, we note that {\Alumina} observed the WD trend only using the G filter, so $\Delta\lambda=0$ and $\Delta n = 0$ (Fig. \ref{fig:pmin-a0-lab-simple}). Moreover, the ND trend is also observed in the three G filter measurements. Therefore, both the WD and ND trends are observed without a change in $n$. As a result, we propose that the effect of $D/\lambda$ is a simpler explanation. However, further experiments may be required to confirm these findings.


\section{Transition to the narrowing and deepening (ND) trend}\label{s:wd-nd transition}
We next consider the lower bound of $D/\lambda$ at which the WD trend stops. We note that experiments involving $D/\lambda \lesssim 1 $ are relatively rare, and the particle size frequency distribution and microscopy images indicate challenges in achieving homogeneous grain sizes in this small regime. As a result, the ND trend itself is not as clear as the WD trend in the literature.

One of the earliest reports documenting the ND trend is shown in Figure 5 of \cite{1969A&A.....2..105D} (also \citealt{1966DollfusMars}). The mixture named ``Adamcik A'', comprising a complex combination of fine-grained goethite, kaolin, hematite, and magnetite, showcases the ND trend when wavelengths increase from $\lambda=0.50\um$ to $0.60\um$: $(\Pmin,\ainv) = (-0.75\%, 24\degr)$ to $(-0.9\%, 20\degr)$. These two points lie closely on the ND trend of $D=0.3\um$ alumina samples in Figs. \ref{fig:pmin-a0-lab-simple} and \ref{fig:pmin-a0-lab}. The authors, however, did not provide further details regarding this behavior due to the inability of the sample to replicate the Martian PPC.

The WD trend among transparent glass powders is shown from $D/\lambda > 100$ to $D/\lambda \sim 1$ (Fig. \ref{fig:pmin-a0-sample}; \citealt{1992Icar...95..283S}). It is not definitively clear if $D/\lambda=1$ represents the WD--ND transition point due to a lack of experiments involving smaller particles and because the transparency of the sample may make the physical processes different from those of regolith material. However, this finding does suggest the possibility that the transition may occur at a $D/\lambda$ value less than $2$.

Specifically, for alumina, Figure 3 in \cite{Shkuratov2002NATO} reports the ND trend for alumina when comparing $D=1\um$ and $D=0.1\um$ samples under $\lambda=0.65\um$, as expected. \cite{2002P&SS...50..849N}, \cite{2018Icar..302..483N} and \cite{2002Icar..159..396S} reported the deepening trend of the NPB as $D/\lambda$ decreases. However, the PPC of the NPB in \cite{2018Icar..302..483N}, spanning from $D=0.1\thru 1.5 \um$, does not show a clear trend in $\ainv$, as expected from the WD and ND trends. This difference may be related to other factors (such as the surface roughening process, grain shape, alignments, imperfect size segregation, and compression, as described in Sect. \ref{ss:effect rough}).

Therefore, the ND trend is not always clearly observed in the literature, as shown in Fig. \ref{fig:pmin-a0-lab-simple}. Combining the observations discussed thus far, we suggest that the WD trend stops (WD--ND transition) at $D/\lambda \sim 1\thru 2$, although this requires further investigation.

\section{A summary of grain size ($D/\lambda$) and albedo effects} \label{s:summ_trends}
\begin{figure*}[tb!]
\centering
\includegraphics[width=0.8\linewidth]{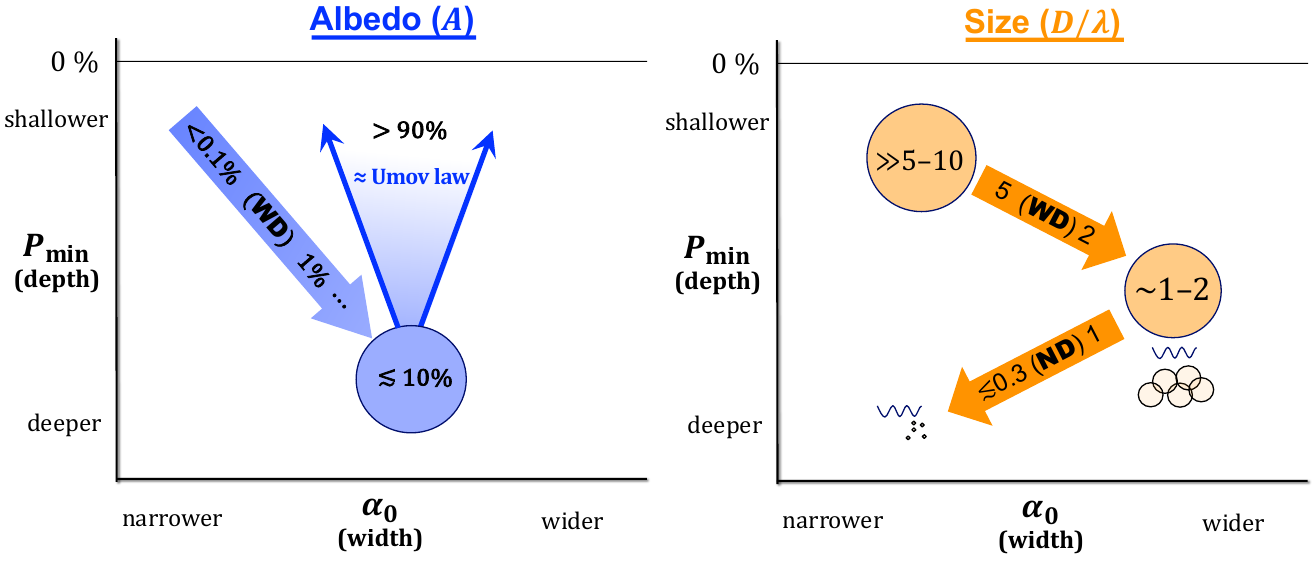}
\caption{Schematic diagram to visualize the trends in albedo and size ($D/\lambda$). The arrows for albedo ($A$; blue) and size ($D/\lambda$; orange) indicate the direction of increasing albedo and decreasing $D/\lambda$ (increasing $\lambda$), respectively.}
\label{fig:trends-schema}
\end{figure*}

Based on existing experimental findings in the academic literature, we examined the patterns within the $\Pmin$--$\ainv$ space. The impacts of the two most important parameters, $D/\lambda$ and albedo, are schematically depicted in Fig. \ref{fig:trends-schema}.

Among these factors, we focus on the importance of the trend by $D/\lambda$ because previous studies separately focused on analyzing the effects of $D$ and $\lambda$. As $\lambda$ increases, given that the albedo does not change significantly,
\begin{enumerate}
\item when the WD trend starts: $D \lesssim (5\thru10)\lambda$,
\item when the WD trend slows down: $D \sim (1\thru2)\lambda$, and
\item when the ND trend appears: $D \lesssim \lambda$.
\end{enumerate}
Based on the discussions and analyses presented throughout this study, it remains uncertain whether the $D$ value derived from this logic corresponds to the most dominant grain size on the surfaces examined. For example, our investigation does not clearly tell whether a small amount of $D=\lambda$ particles in the $D_\mathrm{average}=10\lambda$ sample can induce clear WD and/or ND trends. To address this important question and gain a more comprehensive understanding of these dynamics, future experiments are necessary.

We also note that, the refractive index (and thus the albedo and reflectance) and interparticle distance in terms of wavelength units will change with $\lambda$. In extreme cases, such as those involving ice frost, objects with atmospheres, and near-Sun objects that may emit thermal emission at shorter wavelengths, our straightforward calculation may not be directly applicable. These complications deserve careful consideration and bespoke analyses to provide a more accurate interpretation of polarimetric observations in these contexts. Additionally, the distinction between ``grain size'' and ``microscopic roughness'' is not always sharply defined.

\section{Application to celestial objects}\label{s:application}
Insights into how particle size influences the NPB, as detailed in the preceding sections, can be employed in reverse to estimate particle sizes ($D$) on airless bodies. We tested this phenomenon to validate the feasibility of this methodology. Possible applications for Mercury and Mars are discussed. Finally, a brief discussion of a complementary approach involving thermal modeling is given. Extension to small bodies will be presented in our follow-up paper (\PaperII).

\subsection{Moon}\label{ss:particle size moon}
\subsubsection{Observations}\label{ss:lunar obs}
Although some researchers have mentioned that there is no $\lambda$-dependence in the lunar NPB (e.g., Figure 12 caption of \citealt{Dollfus1971}), others have discussed this phenomenon. Among them, \cite{1964AJ.....69..826G} confirmed the WD trend of the lunar NPB as $\lambda$ increased from the U ($0.36\um$) to the I ($0.94\um$) band. \cite{1965MNRAS.130...83C} reported the same WD trend from filter 94 ($0.459\um$) to 89B ($0.739\um$).

More recently, \cite{1992Icar...95..283S} clearly showed NPB changes over $\lambda$: The NPB is strengthened from blue ($\lambda=0.42 \um$) to red ($0.65 \um$), as shown in Fig. \ref{fig:pmin-a0-sample}. Figure 18 of \cite{1971A&A....10...29D} shows that $\ainv$ changes significantly for the case of the Moon ($\ainv < 21\degr$ at $\lambda=0.327 \um$ to $\ainv \sim 26\degr$ at $\lambda=1.05\um$, and a marginal hint that it remains constant for $\lambda \gtrsim 1\um$). Although they reported that ``neither $\Pmin$ nor $\amin$ change significantly with wavelength,'' the change in $\ainv$ is qualitatively similar to the change in the lunar regions shown in Fig. \ref{fig:pmin-a0-sample}.

\subsubsection{Interpretations}
Because the change in albedo with wavelength is significant for all four types of lunar surfaces, the effect of albedo should be considered. For maria (albedo $\lesssim 10\%$), a WD-like trend is observed when as $\lambda$ increases, aligning with both the albedo trend of darker objects and the effect of decreased $D/\lambda$. For the higher albedo ($>10\%$) regions, however, $\Pmin$ does not change much, while $\ainv$ clearly increases. The albedo effect expects shallowing of the NPB; thus, as for the case of $\mathrm{Fe_2 O_3}$ (Sect. \ref{ss:alb vs size}), we can infer that both the albedo and $D/\lambda$ effects simultaneously and effectively cancel each other except for the increase $\ainv$.

If the $D/\lambda$ effect genuinely impacts the NPB, we can estimate the grain size based on these observations. A WD trend of $D/\lambda$ starting from the U-band to the B-band indicates grains of $D \lesssim (5\thru 10)\lambda \approx 2\thru4 \um$. Moreover, $\ainv $ starts to become constant for $\lambda > 1 \um$ (\citealt{1971A&A....10...29D}; Sect. \ref{ss:lunar obs}), that is, the WD--ND transition. From this, we propose that $D \sim (1\thru2)\lambda \approx 1\thru2 \um$ affects the NPB, which is consistent with the conclusion drawn from the WD trend.

The grain sizes estimated by the two methods agree well with the size-frequency distribution analyses of lunar fine samples \citep{2008ParkJ+Lunar}. This alignment emphasizes that estimating grain size using the initiation of the WD trend is a reliable approach. This further indicates that employing multiwavelength observations of NPB to estimate grain size is a useful approach.

\subsection{Mercury}
Although Mercury was observed at multiple wavelengths \citep{1961plsa.book..343D,1974Icar...23..465D}, the NPB was well covered at only one wavelength, and the PPC near the inversion was available only between $0.520$ and $0.630\um$ at one location. Given the scatter of the data, we cannot find a hint of a change in $\ainv$ (change $ < 1\thru2\degr$). \cite{1974Icar...23..465D} concluded that Mercury has a PPC similar to that of the lightest lunar mare fine, that is, $\Delta \ainv \lesssim 1\degr$ given the wavelength range, according to \cite{1971A&A....10...29D}). Thus, more precise multiwavelength polarimetry in the NPB can shed light on the grain properties of Mercury.

\subsection{Mars}\label{ss:mars}
\begin{figure*}[tb!]
  \centering
  \includegraphics[width=0.7\linewidth]{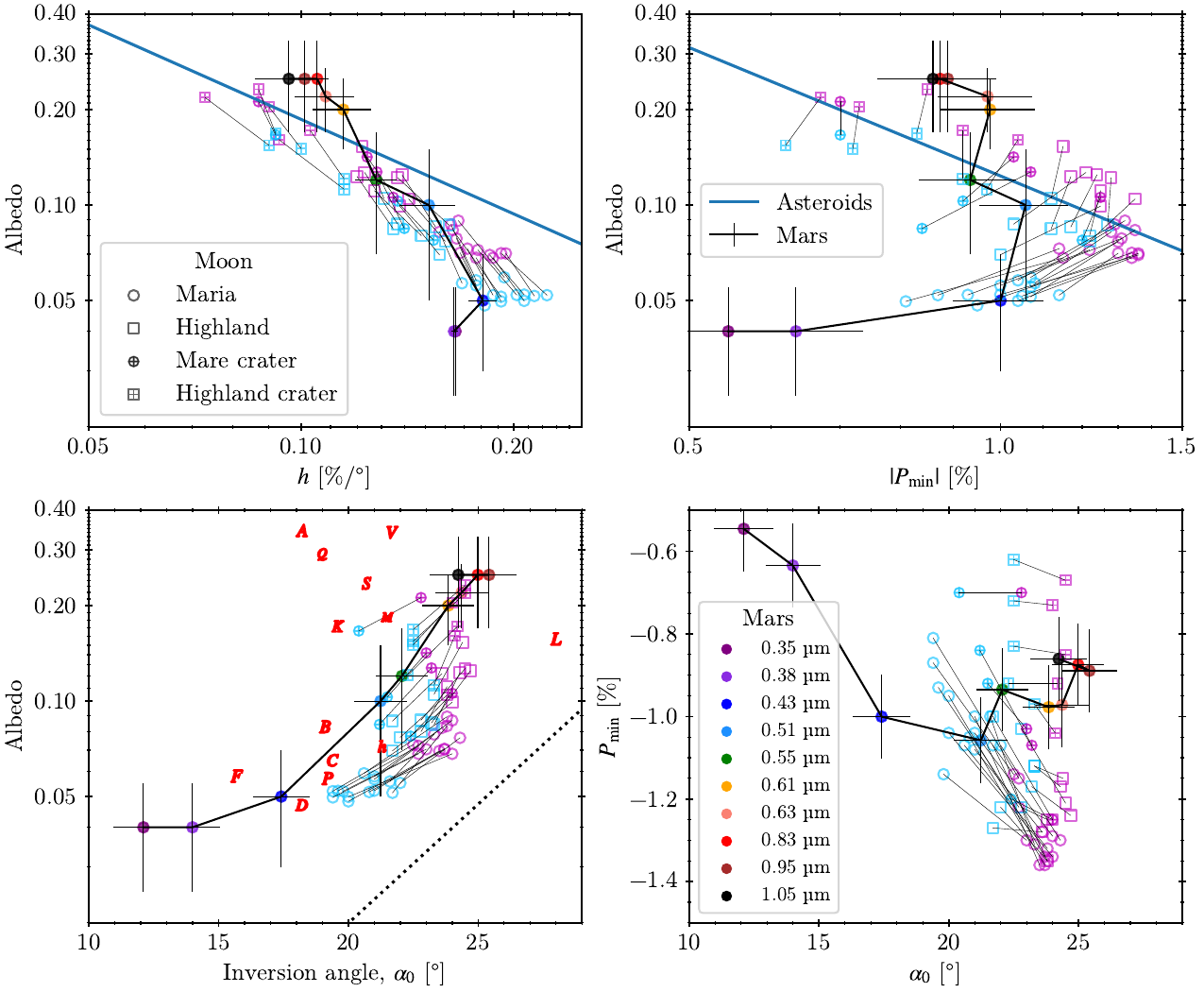}
\caption{Polarimetric parameters of Mars compared to those of the Moon and asteroids. The Mars data are connected in the order of wavelength. The dotted line in the albedo--$\ainv$ space is the same as Fig. \ref{fig:a0-albedo} (Eq. \ref{eq:alb-a0 boundary}).}
  \label{fig:mars}
\end{figure*}

With its thin atmosphere, Mars technically falls outside the scope of our current work. Even so, it is intriguing that multiwavelength polarimetry was conducted on clear and pure surface regions on Mars\footnote{
Called ``R\'{e}gions claires'' in the original publications, which show a minimal effect of atmosphere along the line of sight.
} reveals a definitive change in the NPB (\citealt{1969A&A.....2...63D}; \citealt{1969A&A.....2..105D}; \citealt{1966DollfusMars}). After digitizing their data, we added a data point $(\alpha, \polr) = (0\degr, 0\%)$. Because of the lack of data near $\ainv$ for certain wavelengths, we calculated the polarimetric parameters by two fitting methods: a linear-exponential function and a spline fit. The average of these values was taken, and half of the differences between these two values were regarded as the initial error bars. From visual inspection to the PPCs, $\Delta \Pmin \sim 0.1\pp$ and $\Delta\ainv \sim 1\degr$ are expected (e.g., data points with $\polr < \Pmin-0.2\pp$). We calculated the final error bar by propagating these values as an additional random error. Moreover, from the reflectance spectra \citep{1982JGR....87.3021M, 1994GeoRL..21..353M, 1997Icar..130..449E, 2007Icar..191..581B}, we obtained representative (mean) reflectance values, $r(\lambda)$. The possible ranges (bright and dark regions) at each wavelength are extracted by visual inspection and regarded as error bars. The derived parameters are plotted in four parameter spaces in Fig. \ref{fig:mars}.

The trajectory of Mars in the $\Pmin$--$\ainv$ space displays a typical albedo trend from albedo $<5\%$ to $\sim 30\%$ (Fig. \ref{fig:trends-schema}; Sect. \ref{ss:effect alb}). For longer wavelengths, the albedo remains nearly constant. Thus, if the $D/\lambda$ effect is effective, the NPB must change at the longest wavelengths. Considering that $\ainv$ remains unchanged while $\lambda$ is changed by a factor of $\sim 2$, it is likely that $D/\lambda \gg 10$ is maintained up to $\lambda=1.05\um$, that is, $D\gg 10\um$. Another possible explanation could be the WD--ND transition at $\lambda = 0.6 \thru 1.05 \um$, which translates to $D \sim 0.6\thru1.2 \um$ or $1.05\thru2.1 \um$.

Notably, $D > 10\um$ was obtained based on the $\Pmax$-albedo relationship and thermal modeling for a few selected regions on Mars \citep{1993JGR....98.3413D}. Thus, we prefer the first interpretation that the trace of Mars in the parameter space is driven by the albedo effect, not by the $D/\lambda$ effect, because the grain sizes are too large.

Due to the atmospheric disturbances on Mars, the PPC may not represent the bare surface, even though measurements were deliberately selected based on monitoring the clearest regions. Thus, we refrain from providing further quantitative estimations related to Mars in this context. Instead, we propose that a future near-infrared polarimetry investigation considering regional variation could provide a conclusive perspective on this topic.

\subsection{Thermal modeling: A complementary approach}\label{s:thermal model}
As discussed in Sect. \ref{s:intro}, thermal modeling is also used for indirect measurements of particle sizes. The temperature distribution on an asteroid can be calculated using thermophysical models (TPMs), such as those developed by \cite{1979aste.book..212D} and \cite{1989Icar...78..337S}. This temperature distribution, and the corresponding thermal flux, is closely related to the thermal inertia $\Gamma \coloneqq \sqrt{k \rho c_s}$, thermal conductivity $k$, mass density $\rho$, and specific heat capacity $c_s$.

In the simplest case, a lower $\Gamma$ value indicates less efficient thermal conduction, which can be interpreted as smaller grain sizes in the regolith \citep{2000Icar..148..437M}. The Hayabusa2 mission, which included the MASCOT lander \citep{2013AcAau..91..356T, 2017SSRv..208....3W}, landed on the asteroid (162173) Ryugu. The surface temperature profile from MASCOT's observations revealed that fine grains should not cover surfaces thicker than $50\um$ \citep{2019NatAs...3..971G}. An identical $<50\um$ thickness requirement was found for (101955) Bennu from OSIRIS-REx observations \citep{2020SciA....6.3699R}, and it was concluded that its surface is mostly covered with meter-scale boulders \citep{2019NatAs...3..341D}.

A more realistic model developed by \cite{2013Icar..223..479G} provides a methodology for calculating thermal conductivity by considering the solid spherical particle contact theory and the radiative conduction of opaque particles. This calculation requires assumptions about the properties of the grain, such as Young's modulus, Poisson's ratio, mass density, specific heat capacity, and hemispherical infrared emissivity. These values are assumed and fixed based on the spectral type information of the asteroid. By treating the porosity and grain size as free parameters, the thermal flux of the asteroid can be calculated. Various studies have applied this approach (\citealt{2015Icar..252....1D}, \citealt{2018Icar..309..297H}, \citealt{2015aste.book..107D} and references therein). Recently, \cite{2022PSJ.....3...47M, 2022PSJ.....3..263M} successfully utilized this method to analyze a wide range of asteroids and retrieve information about grain sizes.

This forward modeling approach faces challenges in accurately representing complex reality. One example is the calibration of correction factors, such as the ``$\chi$ factor'' in \cite{2013Icar..223..479G}, which is a multiplication factor to match theory to experiment, based solely on lunar fines data from Apollo 11's 10084-68 \citep{1970GeCAS...1.2045C} and Apollo 12's 12001,19 \citep{1971LPSC....2.2311C}. Another challenge arises from the temperature dependence of properties such as $c_s$ and $k$ \citep{2022PSJ.....3...47M}. Furthermore, the exact determination of thermal inertia through forward modeling relies on data fitting, which is sensitive to observations (bandpass filters, sampling density, etc.) and specific model settings, such as roughness \citep{2014Icar..243...58D} and shape models \citep{2015Icar..256..101H}, which in turn affect particle size determination.

Consequently, while forward modeling offers valuable insights into the underlying physics of observations, it has its own limitations. Thus, alternative independent and empirical approaches such as the polarimetric studies presented in this work hold great value in complementing our understanding. As we have shown, our methodology is sensitive only to particles $D \lesssim 10\lambda$, which is a limitation compared to the thermal modeling approach.
In {\PaperII}, we will discuss our polarimetric observations and analyses, highlighting their strong alignment with grain size estimations derived from an independent methodology by thermal modeling, showcasing the validity of the two approaches. These complementary analyses offer valuable insights, enhancing our understanding of the studied phenomena.

\section{Conclusions and future work}\label{s:conclusion}
In this study, we summarized the effect of physical parameters on the shape of the NPB based on previous experimental and observational studies. The NPB trends are summarized below (also see Fig. \ref{fig:trends-schema}):
\begin{enumerate}
\item With increasing albedo $A$, the NPB exhibited a WD trend until $ A \sim 2\thru10 \%$. The NBP becomes shallower for $A \gtrsim 5\thru 10\%$, which is the expectation of the classical Umov law.
\item With decreasing $D/\lambda$: The NPB stays fixed for $D/\lambda \gg 10$. The WD trend starts at $D/\lambda \sim 5\thru 10$. This trend slows down at $D/\lambda \sim 1\thru 2$, and the ND trend may be observed at $D/\lambda \lesssim \lambda$.
\item The effect of compression (roughness, multiple scattering, and porosity) is case dependent and requires further study.
\item The effect of the real part of the refractive index ($n$) can explain the ND trend, but the $D/\lambda$ effect is a simpler explanation.
\end{enumerate}
In particular, we leveraged the $D/\lambda$ effect to estimate the grain size of selected celestial objects, and the method yielded highly satisfactory results, particularly for the case of the Moon (and Mars). Additionally, we emphasize $D/\lambda$ and that albedo effects can complicate the interpretation of the observed WD trend.

Future laboratory experiments can shed light on the study of regolith on airless bodies. This can include investigating (i) separating the $D/\lambda$ and albedo trends (e.g., using flat reflectance materials), (ii) examining how mixing identical ingredients with different particle sizes influences polarization, (iii) disentangling the contributions of porosity and roughness to the observed polarization, and (iv) designing experiments specifically to quantify the albedo and $D/\lambda$ trends suggested here.

While doing so, maintaining narrow size frequency distributions for each sample can be important, though practically challenging. All other investigations into any parameter space merit science. To ensure the comparability of results across studies, the adoption of a unified protocol for specifying wavelengths and measuring key polarization parameters, such as $\Pmin$, $\amin$, $\ainv$, $h$, and albedo, is paramount (e.g., different definitions of $h$ and albedo complicate direct comparisons). The introduction of
set of widely acceptable parameters related to the skewness of the NPB may shed light on POE. Standardization can enable researchers to draw more robust conclusions and advance our understanding of polarization phenomena.

In the domain of observational science, the study of near-infrared polarimetry methods for asteroids is still in its infancy, with recent studies marking the initial steps in this field (\citealt{2022PSJ.....3...90M, 2023PSJ.....4...93M}; Bach et al. 2024). As noted above, it is crucial to delineate the distinct influences of albedo and $D/\lambda$ effects. To enhance our understanding, we propose a focus on objects where the albedo effect could be separated. We initiated preliminary observations of specific large airless bodies from 2019 to 2021, unaware of the concurrent investigations. In the forthcoming paper of this series (\PaperII), we aim to employ the methodologies established in this work to explore airless objects from an observational viewpoint. This approach will advance our understanding of these cosmic bodies.

\begin{acknowledgements}
This work was supported by a National Research Foundation of Korea (NRF) grant funded by the Korean government (MEST) (No. 2023R1A2C1006180). YPB would like to thank Ang\'{e}lique Pathammavong for a valuable contribution in creating the summary figure. The authors want to thank Joseph Masiero for providing constructive comments as a referee.
\end{acknowledgements}

\bibliographystyle{aa}
\bibliography{ref}

\begin{thebibliography}{120}
\expandafter\ifx\csname natexlab\endcsname\relax\def\natexlab#1{#1}\fi

\bibitem[{{Bach} {et~al.}(2019){Bach}, {Ishiguro}, {Jin}, {Yang}, {Moon},
  {Choi}, {JeongAhn}, {Kim}, \& {Kwak}}]{2019JKAS...52...71B}
{Bach}, Y.~P., {Ishiguro}, M., {Jin}, S., {et~al.} 2019, Journal of Korean
  Astronomical Society, 52, 71

\bibitem[{{Beck} {et~al.}(2021){Beck}, {Schmitt}, {Potin}, {Pommerol}, \&
  {Brissaud}}]{2021Icar..35414066B}
{Beck}, P., {Schmitt}, B., {Potin}, S., {Pommerol}, A., \& {Brissaud}, O. 2021,
  Icarus, 354, 114066

\bibitem[{{Bell} \& {Ansty}(2007)}]{2007Icar..191..581B}
{Bell}, J.~F. \& {Ansty}, T.~M. 2007, Icarus, 191, 581

\bibitem[{{Belskaya} {et~al.}(2015){Belskaya}, {Cellino}, {Gil-Hutton},
  {Muinonen}, \& {Shkuratov}}]{2015aste.book..151B}
{Belskaya}, I., {Cellino}, A., {Gil-Hutton}, R., {Muinonen}, K., \&
  {Shkuratov}, Y. 2015, in Asteroids IV, 151--163

\bibitem[{{Belskaya} {et~al.}(2017){Belskaya}, {Fornasier}, {Tozzi},
  {Gil-Hutton}, {Cellino}, {Antonyuk}, {Krugly}, {Dovgopol}, \&
  {Faggi}}]{2017Icar..284...30B}
{Belskaya}, I.~N., {Fornasier}, S., {Tozzi}, G.~P., {et~al.} 2017, Icarus, 284,
  30

\bibitem[{{Belskaya} {et~al.}(2005){Belskaya}, {Shkuratov}, {Efimov},
  {Shakhovskoy}, {Gil-Hutton}, {Cellino}, {Zubko}, {Ovcharenko}, {Bondarenko},
  {Shevchenko}, {Fornasier}, \& {Barbieri}}]{2005Icar..178..213B}
{Belskaya}, I.~N., {Shkuratov}, Y.~G., {Efimov}, Y.~S., {et~al.} 2005, Icarus,
  178, 213

\bibitem[{{Bohren} \& {Huffman}(1983)}]{1983asls.book.....B}
{Bohren}, C.~F. \& {Huffman}, D.~R. 1983, {Absorption and scattering of light
  by small particles}

\bibitem[{{Bowell} {et~al.}(1973){Bowell}, {Dollfus}, {Zellner}, \&
  {Geake}}]{1973LPSC....4.3167B}
{Bowell}, E., {Dollfus}, A., {Zellner}, B., \& {Geake}, J.~E. 1973, Lunar and
  Planetary Science Conference Proceedings, 4, 3167

\bibitem[{{Brewster}(1863)}]{Brewster1861}
{Brewster}, D. 1863, Transactions of the Royal Society of Edinburgh, 23, 205

\bibitem[{{Cellino} {et~al.}(2015){Cellino}, {Bagnulo}, {Gil-Hutton}, {Tanga},
  {Ca{\~n}ada-Assandri}, \& {Tedesco}}]{2015MNRAS.451.3473C}
{Cellino}, A., {Bagnulo}, S., {Gil-Hutton}, R., {et~al.} 2015, Monthly Notices
  of the Royal Astronomical Society, 451, 3473

\bibitem[{{Clarke}(1965)}]{1965MNRAS.130...83C}
{Clarke}, D. 1965, Monthly Notices of the Royal Astronomical Society, 130, 83

\bibitem[{{Cremers} \& {Birkebak}(1971)}]{1971LPSC....2.2311C}
{Cremers}, C.~J. \& {Birkebak}, R.~C. 1971, Lunar and Planetary Science
  Conference Proceedings, 2, 2311

\bibitem[{{Cremers} {et~al.}(1970){Cremers}, {Birkebak}, \&
  {Dawson}}]{1970GeCAS...1.2045C}
{Cremers}, C.~J., {Birkebak}, R.~C., \& {Dawson}, J.~P. 1970, Geochimica et
  Cosmochimica Acta Supplement, 1, 2045

\bibitem[{{Dabrowska} {et~al.}(2015){Dabrowska}, {Mu{\~n}oz}, {Moreno},
  {Ramos}, {Mart{\'\i}nez-Fr{\'\i}as}, \& {Wurm}}]{2015Icar..250...83D}
{Dabrowska}, D.~D., {Mu{\~n}oz}, O., {Moreno}, F., {et~al.} 2015, Icarus, 250,
  83

\bibitem[{{Daly} {et~al.}(2023){Daly}, {Ernst}, {Barnouin}, {Chabot}, {Rivkin},
  {Cheng}, {Adams}, {Agrusa}, {Abel}, {Alford}, {Asphaug}, {Atchison},
  {Badger}, {Baki}, {Ballouz}, {Bekker}, {Bellerose}, {Bhaskaran}, {Buratti},
  {Cambioni}, {Chen}, {Chesley}, {Chiu}, {Collins}, {Cox}, {DeCoster},
  {Ericksen}, {Espiritu}, {Faber}, {Farnham}, {Ferrari}, {Fletcher}, {Gaskell},
  {Graninger}, {Haque}, {Harrington-Duff}, {Hefter}, {Herreros}, {Hirabayashi},
  {Huang}, {Hsieh}, {Jacobson}, {Jenkins}, {Jensenius}, {John}, {Jutzi},
  {Kohout}, {Krueger}, {Laipert}, {Lopez}, {Luther}, {Lucchetti}, {Mages},
  {Marchi}, {Martin}, {McQuaide}, {Michel}, {Moskovitz}, {Murphy}, {Murdoch},
  {Naidu}, {Nair}, {Nolan}, {Orm{\"o}}, {Pajola}, {Palmer}, {Peachey},
  {Pravec}, {Raducan}, {Ramesh}, {Ramirez}, {Reynolds}, {Richman}, {Robin},
  {Rodriguez}, {Roufberg}, {Rush}, {Sawyer}, {Scheeres}, {Scheirich},
  {Schwartz}, {Shannon}, {Shapiro}, {Shearer}, {Smith}, {Steele}, {Steckloff},
  {Stickle}, {Sunshine}, {Superfin}, {Tarzi}, {Thomas}, {Thomas},
  {Trigo-Rodr{\'\i}guez}, {Tropf}, {Vaughan}, {Velez}, {Waller}, {Wilson},
  {Wortman}, \& {Zhang}}]{2023Natur.616..443D}
{Daly}, R.~T., {Ernst}, C.~M., {Barnouin}, O.~S., {et~al.} 2023, Nature, 616,
  443

\bibitem[{{Davidsson} \& {Rickman}(2014)}]{2014Icar..243...58D}
{Davidsson}, B. J.~R. \& {Rickman}, H. 2014, Icarus, 243, 58

\bibitem[{{Davidsson} {et~al.}(2015){Davidsson}, {Rickman}, {Bandfield},
  {Groussin}, {Guti{\'e}rrez}, {Wilska}, {Capria}, {Emery}, {Helbert}, {Jorda},
  {Maturilli}, \& {Mueller}}]{2015Icar..252....1D}
{Davidsson}, B. J.~R., {Rickman}, H., {Bandfield}, J.~L., {et~al.} 2015,
  Icarus, 252, 1

\bibitem[{{Delbo} {et~al.}(2015){Delbo}, {Mueller}, {Emery}, {Rozitis}, \&
  {Capria}}]{2015aste.book..107D}
{Delbo}, M., {Mueller}, M., {Emery}, J.~P., {Rozitis}, B., \& {Capria}, M.~T.
  2015, in Asteroids IV, 107--128

\bibitem[{{Dellagiustina} {et~al.}(2019){Dellagiustina}, {Emery}, {Golish},
  {Rozitis}, {Bennett}, {Burke}, {Ballouz}, {Becker}, {Christensen}, {Drouet
  D'Aubigny}, {Hamilton}, {Reuter}, {Rizk}, {Simon}, {Asphaug}, {Bandfield},
  {Barnouin}, {Barucci}, {Bierhaus}, {Binzel}, {Bottke}, {Bowles}, {Campins},
  {Clark}, {Clark}, {Connolly}, {Daly}, {Leon}, {Delbo'}, {Deshapriya},
  {Elder}, {Fornasier}, {Hergenrother}, {Howell}, {Jawin}, {Kaplan}, {Kareta},
  {Le Corre}, {Li}, {Licandro}, {Lim}, {Michel}, {Molaro}, {Nolan}, {Pajola},
  {Popescu}, {Garcia}, {Ryan}, {Schwartz}, {Shultz}, {Siegler}, {Smith},
  {Tatsumi}, {Thomas}, {Walsh}, {Wolner}, {Zou}, {Lauretta}, \& {Osiris-Rex
  Team}}]{2019NatAs...3..341D}
{Dellagiustina}, D.~N., {Emery}, J.~P., {Golish}, D.~R., {et~al.} 2019, Nature
  Astronomy, 3, 341

\bibitem[{{Devog{\`e}le} {et~al.}(2018){Devog{\`e}le}, {Cellino}, {Borisov},
  {Bendjoya}, {Rivet}, {Abe}, {Bagnulo}, {Christou}, {Vernet}, {Donchev},
  {Belskaya}, {Bonev}, \& {Krugly}}]{2018MNRAS.479.3498D}
{Devog{\`e}le}, M., {Cellino}, A., {Borisov}, G., {et~al.} 2018, Monthly
  Notices of the Royal Astronomical Society, 479, 3498

\bibitem[{{Dickel}(1979)}]{1979aste.book..212D}
{Dickel}, J.~R. 1979, in Asteroids, ed. T.~{Gehrels} \& M.~S. {Matthews},
  212--221

\bibitem[{Dodge(1986)}]{1986_Dodge}
Dodge, M.~J. 1986, in CRC handbook of laser science and technology. Volume 4.
  Optical materials, Part 2 - Properties, ed. M.~J. Weber (CRC Press), 21

\bibitem[{{Dollfus}(1956)}]{1956AnAp...19...83D}
{Dollfus}, A. 1956, Annales d'Astrophysique, 19, 83

\bibitem[{{Dollfus}(1961)}]{1961plsa.book..343D}
{Dollfus}, A. 1961, in Planets and Satellites, ed. G.~P. {Kuiper} \& B.~M.
  {Middlehurst}, 343

\bibitem[{Dollfus(1971)}]{Dollfus1971}
Dollfus, A. 1971, in {Physical Studies of Minor Planets}, ed. {Gehrels, Tom}
  (ational Aeronautics and Space Administration, Washinton, D. C.), 95--116

\bibitem[{{Dollfus} \& {Auriere}(1974)}]{1974Icar...23..465D}
{Dollfus}, A. \& {Auriere}, M. 1974, Icarus, 23, 465

\bibitem[{{Dollfus} \& {Bowell}(1971)}]{1971A&A....10...29D}
{Dollfus}, A. \& {Bowell}, E. 1971, Astronomy and Astrophysics, 10, 29

\bibitem[{{Dollfus} {et~al.}(1971{\natexlab{a}}){Dollfus}, {Bowell}, \&
  {Titulaer}}]{1971A&A....10..450D}
{Dollfus}, A., {Bowell}, E., \& {Titulaer}, C. 1971{\natexlab{a}}, Astronomy
  and Astrophysics, 10, 450

\bibitem[{{Dollfus} \& {Deschamps}(1986)}]{1986Icar...67...37D}
{Dollfus}, A. \& {Deschamps}, M. 1986, Icarus, 67, 37

\bibitem[{{Dollfus} {et~al.}(1993){Dollfus}, {Deschamps}, \&
  {Zimbelman}}]{1993JGR....98.3413D}
{Dollfus}, A., {Deschamps}, M., \& {Zimbelman}, J.~R. 1993, Journal of
  Geophysics Research, 98, 3413

\bibitem[{{Dollfus} \& {Focas}(1969)}]{1969A&A.....2...63D}
{Dollfus}, A. \& {Focas}, J. 1969, Astronomy and Astrophysics, 2, 63

\bibitem[{{Dollfus} {et~al.}(1969){Dollfus}, {Focas}, \&
  {Bowell}}]{1969A&A.....2..105D}
{Dollfus}, A., {Focas}, J., \& {Bowell}, E. 1969, Astronomy and Astrophysics,
  2, 105

\bibitem[{{Dollfus} \& {Focas}(1966)}]{1966DollfusMars}
{Dollfus}, A. \& {Focas}, J.~H. 1966, Polarimetric study of the planet Mars (US
  Air Force Cambridge Research Laboratories)

\bibitem[{{Dollfus} \& {Geake}(1975)}]{1975LPSC....6.2749D}
{Dollfus}, A. \& {Geake}, J.~E. 1975, Lunar and Planetary Science Conference
  Proceedings, 3, 2749

\bibitem[{{Dollfus} \& {Geake}(1977)}]{1977RSPTA.285..397D}
{Dollfus}, A. \& {Geake}, J.~E. 1977, Philosophical Transactions of the Royal
  Society of London Series A, 285, 397

\bibitem[{{Dollfus} {et~al.}(1971{\natexlab{b}}){Dollfus}, {Geake}, \&
  {Titulaer}}]{1971LPSC....2.2285D}
{Dollfus}, A., {Geake}, J.~E., \& {Titulaer}, C. 1971{\natexlab{b}}, Lunar and
  Planetary Science Conference Proceedings, 2, 2285

\bibitem[{{Dollfus} {et~al.}(1989){Dollfus}, {Wolff}, {Geake}, {Lupishko}, \&
  {Dougherty}}]{1989aste.conf..594D}
{Dollfus}, A., {Wolff}, M., {Geake}, J.~E., {Lupishko}, D.~F., \& {Dougherty},
  L.~M. 1989, in Asteroids II, ed. R.~P. {Binzel}, T.~{Gehrels}, \& M.~S.
  {Matthews}, 594--616

\bibitem[{{Dollfus} \& {Zellner}(1979)}]{1979aste.book..170D}
{Dollfus}, A. \& {Zellner}, B. 1979, in Asteroids, ed. T.~{Gehrels} \& M.~S.
  {Matthews}, 170--183

\bibitem[{{Dollfus}(1955)}]{1955PhDT........37D}
{Dollfus}, A.~C. 1955, PhD thesis, Universite Pierre et Marie Curie (Paris VI),
  France

\bibitem[{{Dougherty} \& {Geake}(1994)}]{1994MNRAS.271..343D}
{Dougherty}, L.~M. \& {Geake}, J.~E. 1994, Monthly Notices of the Royal
  Astronomical Society, 271, 343

\bibitem[{{Erard} \& {Calvin}(1997)}]{1997Icar..130..449E}
{Erard}, S. \& {Calvin}, W. 1997, Icarus, 130, 449

\bibitem[{{Escobar-Cerezo} {et~al.}(2018){Escobar-Cerezo}, {Mu{\~n}oz},
  {Moreno}, {Guirado}, {G{\'o}mez Mart{\'\i}n}, {Goguen}, {Garboczi},
  {Chiaramonti}, {Lafarge}, \& {West}}]{2018ApJS..235...19E}
{Escobar-Cerezo}, J., {Mu{\~n}oz}, O., {Moreno}, F., {et~al.} 2018,
  Astrophysical Journal, Supplement, 235, 19

\bibitem[{{Frattin} {et~al.}(2022){Frattin}, {Martikainen}, {Mu{\~n}oz},
  {G{\'o}mez-Mart{\'\i}n}, {Jardiel}, {Cellino}, {Libourel}, {Muinonen},
  {Peiteado}, \& {Tanga}}]{2022MNRAS.517.5463F}
{Frattin}, E., {Martikainen}, J., {Mu{\~n}oz}, O., {et~al.} 2022, Monthly
  Notices of the Royal Astronomical Society, 517, 5463

\bibitem[{{Geake} \& {Dollfus}(1986)}]{1986MNRAS.218...75G}
{Geake}, J.~E. \& {Dollfus}, A. 1986, Monthly Notices of the Royal Astronomical
  Society, 218, 75

\bibitem[{{Geake} {et~al.}(1970){Geake}, {Dollfus}, {Garlick}, {Lamb},
  {Walker}, {Steigmann}, \& {Titulaer}}]{1970GeCAS...1.2127G}
{Geake}, J.~E., {Dollfus}, A., {Garlick}, G.~F.~J., {et~al.} 1970, Geochimica
  et Cosmochimica Acta Supplement, 1, 2127

\bibitem[{{Geake} \& {Geake}(1990)}]{1990MNRAS.245...46G}
{Geake}, J.~E. \& {Geake}, M. 1990, Monthly Notices of the Royal Astronomical
  Society, 245, 46

\bibitem[{{Geake} {et~al.}(1984){Geake}, {Geake}, \&
  {Zellner}}]{1984MNRAS.210...89G}
{Geake}, J.~E., {Geake}, M., \& {Zellner}, B.~H. 1984, Monthly Notices of the
  Royal Astronomical Society, 210, 89

\bibitem[{{Geem} {et~al.}(2023){Geem}, {Ishiguro}, {Granvik}, {Naito},
  {Akitaya}, {Sekiguchi}, {Hasegawa}, {Kuroda}, {Oono}, {Bach}, {Jin},
  {Imazawa}, {Kawabata}, {Takagi}, {Yoshikawa}, {Djupvik}, {Gadeberg},
  {Pursimo}, {Pedros}, {Thomsen}, \& {Gray}}]{2023MNRAS.525L..17G}
{Geem}, J., {Ishiguro}, M., {Granvik}, M., {et~al.} 2023, Monthly Notices of
  the Royal Astronomical Society, 525, L17

\bibitem[{{Geem} {et~al.}(2022){Geem}, {Ishiguro}, {Takahashi}, {Akitaya},
  {Kawabata}, {Nakaoka}, {Imazawa}, {Mori}, {Jin}, {Bach}, {Jo}, {Kuroda},
  {Hasegawa}, {Yoshida}, {Ishibashi}, {Sekiguchi}, {Beniyama}, {Arai}, {Ikeda},
  {Shinnaka}, {Granvik}, {Siltala}, {Djupvik}, {Kasikov}, {Pinter}, \&
  {Knudstrup}}]{2022MNRAS.516L..53G}
{Geem}, J., {Ishiguro}, M., {Takahashi}, J., {et~al.} 2022, Monthly Notices of
  the Royal Astronomical Society, 516, L53

\bibitem[{{Gehrels} {et~al.}(1964){Gehrels}, {Coffeen}, \&
  {Owings}}]{1964AJ.....69..826G}
{Gehrels}, T., {Coffeen}, T., \& {Owings}, D. 1964, Astronomical Journal, 69,
  826

\bibitem[{Gil-Hutton(2017)}]{gil-hutton-data}
Gil-Hutton, R. 2017, in ACM (Asteroid, Comets, Meteors) 2017, Montevideo,
  Poster1.d.17

\bibitem[{{Grott} {et~al.}(2019){Grott}, {Knollenberg}, {Hamm}, {Ogawa},
  {Jaumann}, {Otto}, {Delbo}, {Michel}, {Biele}, {Neumann}, {Knapmeyer},
  {K{\"u}hrt}, {Senshu}, {Okada}, {Helbert}, {Maturilli}, {M{\"u}ller},
  {Hagermann}, {Sakatani}, {Tanaka}, {Arai}, {Mottola}, {Tachibana}, {Pelivan},
  {Drube}, {Vincent}, {Yano}, {Pilorget}, {Matz}, {Schmitz}, {Koncz},
  {Schr{\"o}der}, {Trauthan}, {Schlotterer}, {Krause}, {Ho}, \&
  {Moussi-Soffys}}]{2019NatAs...3..971G}
{Grott}, M., {Knollenberg}, J., {Hamm}, M., {et~al.} 2019, Nature Astronomy, 3,
  971

\bibitem[{{Gundlach} \& {Blum}(2013)}]{2013Icar..223..479G}
{Gundlach}, B. \& {Blum}, J. 2013, Icarus, 223, 479

\bibitem[{{Hadamcik} {et~al.}(2023){Hadamcik}, {Renard}, {Lasue},
  {Levasseur-Regourd}, \& {Ishiguro}}]{2023MNRAS.520.1963H}
{Hadamcik}, E., {Renard}, J.~B., {Lasue}, J., {Levasseur-Regourd}, A.~C., \&
  {Ishiguro}, M. 2023, Monthly Notices of the Royal Astronomical Society, 520,
  1963

\bibitem[{{Hadamcik} {et~al.}(2009){Hadamcik}, {Renard}, {Levasseur-Regourd},
  {Lasue}, {Alcouffe}, \& {Francis}}]{2009JQSRT.110.1755H}
{Hadamcik}, E., {Renard}, J.~B., {Levasseur-Regourd}, A.~C., {et~al.} 2009,
  Journal of Quantitiative Spectroscopy and Radiative Transfer, 110, 1755

\bibitem[{{Hanu{\v{s}}} {et~al.}(2015){Hanu{\v{s}}}, {Delbo'}, {{\v{D}}urech},
  \& {Al{\'\i}-Lagoa}}]{2015Icar..256..101H}
{Hanu{\v{s}}}, J., {Delbo'}, M., {{\v{D}}urech}, J., \& {Al{\'\i}-Lagoa}, V.
  2015, Icarus, 256, 101

\bibitem[{{Hanu{\v{s}}} {et~al.}(2018){Hanu{\v{s}}}, {Delbo'}, {{\v{D}}urech},
  \& {Al{\'\i}-Lagoa}}]{2018Icar..309..297H}
{Hanu{\v{s}}}, J., {Delbo'}, M., {{\v{D}}urech}, J., \& {Al{\'\i}-Lagoa}, V.
  2018, Icarus, 309, 297

\bibitem[{Hapke(2012)}]{Hapke2012}
Hapke, B. 2012, Theory of reflectance and emittance spectroscopy (Cambridge
  university press)

\bibitem[{{Hosseini}(2008)}]{2008PSSBR.245.2800H}
{Hosseini}, S.~M. 2008, Physica Status Solidi B Basic Research, 245, 2800

\bibitem[{{Ishiguro} {et~al.}(2022){Ishiguro}, {Bach}, {Geem}, {Naito},
  {Kuroda}, {Im}, {Lee}, {Seo}, {Jin}, {Kwon}, {Oono}, {Takagi}, {Sato},
  {Kuramoto}, {Ito}, {Hasegawa}, {Yoshida}, {Arai}, {Akitaya}, {Sekiguchi},
  {Okazaki}, {Imai}, {Ohtsuka}, {Watanabe}, {Takahashi}, {Devog{\`e}le},
  {Fedorets}, {Siltala}, \& {Granvik}}]{2022MNRAS.509.4128I}
{Ishiguro}, M., {Bach}, Y.~P., {Geem}, J., {et~al.} 2022, Monthly Notices of
  the Royal Astronomical Society, 509, 4128

\bibitem[{{Ishiguro} {et~al.}(2017){Ishiguro}, {Kuroda}, {Watanabe}, {Bach},
  {Kim}, {Lee}, {Sekiguchi}, {Naito}, {Ohtsuka}, {Hanayama}, {Hasegawa},
  {Usui}, {Urakawa}, {Imai}, {Sato}, \& {Kuramoto}}]{2017AJ....154..180I}
{Ishiguro}, M., {Kuroda}, D., {Watanabe}, M., {et~al.} 2017, Astronomical
  Journal, 154, 180

\bibitem[{{Ito} {et~al.}(2018){Ito}, {Ishiguro}, {Arai}, {Imai}, {Sekiguchi},
  {Bach}, {Kwon}, {Kobayashi}, {Ishimaru}, {Naito}, {Watanabe}, \&
  {Kuramoto}}]{2018NatCo...9.2486I}
{Ito}, T., {Ishiguro}, M., {Arai}, T., {et~al.} 2018, Nature Communications, 9,
  2486

\bibitem[{{Kenknight} {et~al.}(1967){Kenknight}, {Rosenberg}, \&
  {Wehner}}]{1967JGR....72.3105K}
{Kenknight}, C.~E., {Rosenberg}, D.~L., \& {Wehner}, G.~K. 1967, Journal of
  Geophysics Research, 72, 3105

\bibitem[{{Le Bertre} \& {Zellner}(1980)}]{1980Icar...43..172L}
{Le Bertre}, T. \& {Zellner}, B. 1980, Icarus, 43, 172

\bibitem[{{Levasseur-Regourd} {et~al.}(2015){Levasseur-Regourd}, {Renard},
  {Shkuratov}, \& {Hadamcik}}]{2015psps.book...62L}
{Levasseur-Regourd}, A.-C., {Renard}, J.-B., {Shkuratov}, Y., \& {Hadamcik}, E.
  2015, in Polarimetry of Stars and Planetary Systems, 35

\bibitem[{{Lupishko}(2018)}]{2018SoSyR..52...98L}
{Lupishko}, D.~F. 2018, Solar System Research, 52, 98

\bibitem[{{Lupishko} \& {Belskaya}(1989)}]{1989Icar...78..395L}
{Lupishko}, D.~F. \& {Belskaya}, I.~N. 1989, Icarus, 78, 395

\bibitem[{{Lyot}(1929)}]{1929PhDT.........9L}
{Lyot}, B. 1929, PhD thesis, Universite Pierre et Marie Curie (Paris VI),
  France

\bibitem[{{MacLennan} \& {Emery}(2022{\natexlab{a}})}]{2022PSJ.....3..263M}
{MacLennan}, E.~M. \& {Emery}, J.~P. 2022{\natexlab{a}}, Planetary Science
  Journal, 3, 263

\bibitem[{{MacLennan} \& {Emery}(2022{\natexlab{b}})}]{2022PSJ.....3...47M}
{MacLennan}, E.~M. \& {Emery}, J.~P. 2022{\natexlab{b}}, Planetary Science
  Journal, 3, 47

\bibitem[{{Malitson}(1962)}]{1962JOSA...52.1377M}
{Malitson}, I.~H. 1962, Journal of the Optical Society of America (1917-1983),
  52, 1377

\bibitem[{Malitson \& Dodge(1972)}]{1972_Malitson}
Malitson, I.~H. \& Dodge, M.~J. 1972, Journal of the Optical Society of
  America, 62, 1405

\bibitem[{{Masiero} {et~al.}(2009){Masiero}, {Hartzell}, \&
  {Scheeres}}]{2009AJ....138.1557M}
{Masiero}, J., {Hartzell}, C., \& {Scheeres}, D.~J. 2009, Astronomical Journal,
  138, 1557

\bibitem[{{Masiero} {et~al.}(2023){Masiero}, {Devog{\`e}le}, {Macias},
  {Castaneda Jaimes}, \& {Cellino}}]{2023PSJ.....4...93M}
{Masiero}, J.~R., {Devog{\`e}le}, M., {Macias}, I., {Castaneda Jaimes}, J., \&
  {Cellino}, A. 2023, Planetary Science Journal, 4, 93

\bibitem[{{Masiero} {et~al.}(2022){Masiero}, {Tinyanont}, \&
  {Millar-Blanchaer}}]{2022PSJ.....3...90M}
{Masiero}, J.~R., {Tinyanont}, S., \& {Millar-Blanchaer}, M.~A. 2022, Planetary
  Science Journal, 3, 90

\bibitem[{{McCord} {et~al.}(1982){McCord}, {Clark}, \&
  {Singer}}]{1982JGR....87.3021M}
{McCord}, T.~B., {Clark}, R.~N., \& {Singer}, R.~B. 1982, Journal of Geophysics
  Research, 87, 3021

\bibitem[{{Mellon} {et~al.}(2000){Mellon}, {Jakosky}, {Kieffer}, \&
  {Christensen}}]{2000Icar..148..437M}
{Mellon}, M.~T., {Jakosky}, B.~M., {Kieffer}, H.~H., \& {Christensen}, P.~R.
  2000, Icarus, 148, 437

\bibitem[{{Mu{\~n}oz} {et~al.}(2021){Mu{\~n}oz}, {Frattin}, {Jardiel},
  {G{\'o}mez-Mart{\'\i}n}, {Moreno}, {Ramos}, {Guirado}, {Peiteado},
  {Caballero}, {Milli}, \& {M{\'e}nard}}]{2021ApJS..256...17M}
{Mu{\~n}oz}, O., {Frattin}, E., {Jardiel}, T., {et~al.} 2021, Astrophysical
  Journal, Supplement, 256, 17

\bibitem[{{Muinonen} {et~al.}(2002){Muinonen}, {Piironen}, {Shkuratov},
  {Ovcharenko}, \& {Clark}}]{2002aste.book..123M}
{Muinonen}, K., {Piironen}, J., {Shkuratov}, Y.~G., {Ovcharenko}, A., \&
  {Clark}, B.~E. 2002, in Asteroids III, 123--138

\bibitem[{Muinonen {et~al.}(2002)Muinonen, Videen, Zubko, \&
  Shkuratov}]{Muinonen2002NATO}
Muinonen, K., Videen, G., Zubko, E., \& Shkuratov, Y. 2002, Numerical
  Techniques for Backscattering by Random Media, ed. G.~Videen \& M.~Kocifaj
  (Dordrecht: Springer Netherlands), 261--282

\bibitem[{{Muinonen} {et~al.}(2002){Muinonen}, {Videen}, {Zubko}, \&
  {Shkuratov}}]{2002ocd..conf..261M}
{Muinonen}, K., {Videen}, G., {Zubko}, E., \& {Shkuratov}, Y. 2002, in Optics
  of Cosmic Dust, ed. G.~{Videen} \& M.~{Kocifaj}, 261

\bibitem[{{Muinonen}(1990)}]{1990PhDT.......329M}
{Muinonen}, K.~O. 1990, PhD thesis, University of Helsinki, Finland

\bibitem[{{Mustard} \& {Bell}(1994)}]{1994GeoRL..21..353M}
{Mustard}, J.~F. \& {Bell}, J.~F. 1994, Geophysics Research Letters, 21, 353

\bibitem[{{Nelson} {et~al.}(2018){Nelson}, {Boryta}, {Hapke}, {Manatt},
  {Shkuratov}, {Psarev}, {Vandervoort}, {Kroner}, {Nebedum}, {Vides}, \&
  {Qui{\~n}ones}}]{2018Icar..302..483N}
{Nelson}, R.~M., {Boryta}, M.~D., {Hapke}, B.~W., {et~al.} 2018, Icarus, 302,
  483

\bibitem[{{Nelson} {et~al.}(2002){Nelson}, {Smythe}, {Hapke}, \&
  {Hale}}]{2002P&SS...50..849N}
{Nelson}, R.~M., {Smythe}, W.~D., {Hapke}, B.~W., \& {Hale}, A.~S. 2002,
  Planetary Space Science, 50, 849

\bibitem[{{Ovcharenko} {et~al.}(2006){Ovcharenko}, {Bondarenko}, {Zubko},
  {Shkuratov}, {Videen}, {Nelson}, \& {Smythe}}]{2006JQSRT.101..394O}
{Ovcharenko}, A.~A., {Bondarenko}, S.~Y., {Zubko}, E.~S., {et~al.} 2006,
  Journal of Quantitiative Spectroscopy and Radiative Transfer, 101, 394

\bibitem[{Park {et~al.}(2008)Park, Liu, Kihm, \& Taylor}]{2008ParkJ+Lunar}
Park, J., Liu, Y., Kihm, K.~D., \& Taylor, L.~A. 2008, Journal of Aerospace
  Engineering, 21, 266

\bibitem[{{Poch} {et~al.}(2018){Poch}, {Cerubini}, {Pommerol}, {Jost}, \&
  {Thomas}}]{2018JGRE..123.2564P}
{Poch}, O., {Cerubini}, R., {Pommerol}, A., {Jost}, B., \& {Thomas}, N. 2018,
  Journal of Geophysical Research (Planets), 123, 2564

\bibitem[{{Querry}(1985)}]{1985umo..rept.....Q}
{Querry}, M.~R. 1985, {Optical constants}, Contractor Report, Sep. 1982 - May
  1984 Missouri Univ., Kansas City.

\bibitem[{{Rozitis} {et~al.}(2020){Rozitis}, {Ryan}, {Emery}, {Christensen},
  {Hamilton}, {Simon}, {Reuter}, {Al Asad}, {Ballouz}, {Bandfield}, {Barnouin},
  {Bennett}, {Bernacki}, {Burke}, {Cambioni}, {Clark}, {Daly}, {Delbo},
  {DellaGiustina}, {Elder}, {Hanna}, {Haberle}, {Howell}, {Golish}, {Jawin},
  {Kaplan}, {Lim}, {Molaro}, {Munoz}, {Nolan}, {Rizk}, {Siegler}, {Susorney},
  {Walsh}, \& {Lauretta}}]{2020SciA....6.3699R}
{Rozitis}, B., {Ryan}, A.~J., {Emery}, J.~P., {et~al.} 2020, Science Advances,
  6, eabc3699

\bibitem[{{Shkuratov} {et~al.}(1992){Shkuratov}, {Opanasenko}, \&
  {Kreslavskii}}]{1992Icar...95..283S}
{Shkuratov}, I.~G., {Opanasenko}, N.~V., \& {Kreslavskii}, M.~A. 1992, Icarus,
  95, 283

\bibitem[{{Shkuratov} {et~al.}(2007){Shkuratov}, {Bondarenko}, {Kaydash},
  {Videen}, {Mu{\~n}oz}, \& {Volten}}]{2007JQSRT.106..487S}
{Shkuratov}, Y., {Bondarenko}, S., {Kaydash}, V., {et~al.} 2007, Journal of
  Quantitiative Spectroscopy and Radiative Transfer, 106, 487

\bibitem[{{Shkuratov} {et~al.}(2006){Shkuratov}, {Bondarenko}, {Ovcharenko},
  {Pieters}, {Hiroi}, {Volten}, {Mu{\~n}oz}, \& {Videen}}]{2006JQSRT.100..340S}
{Shkuratov}, Y., {Bondarenko}, S., {Ovcharenko}, A., {et~al.} 2006, Journal of
  Quantitiative Spectroscopy and Radiative Transfer, 100, 340

\bibitem[{{Shkuratov} {et~al.}(2002){Shkuratov}, {Ovcharenko}, {Zubko},
  {Miloslavskaya}, {Muinonen}, {Piironen}, {Nelson}, {Smythe}, {Rosenbush}, \&
  {Helfenstein}}]{2002Icar..159..396S}
{Shkuratov}, Y., {Ovcharenko}, A., {Zubko}, E., {et~al.} 2002, Icarus, 159, 396

\bibitem[{{Shkuratov} {et~al.}(2004){Shkuratov}, {Ovcharenko}, {Zubko},
  {Volten}, {Munoz}, \& {Videen}}]{2004JQSRT..88..267S}
{Shkuratov}, Y., {Ovcharenko}, A., {Zubko}, E., {et~al.} 2004, Journal of
  Quantitiative Spectroscopy and Radiative Transfer, 88, 267

\bibitem[{{Shkuratov}(1987)}]{1987SvAL...13..182S}
{Shkuratov}, Y.~G. 1987, Soviet Astronomy Letters, 13, 182

\bibitem[{{Shkuratov}(1989)}]{1989SoSyR..23..111S}
{Shkuratov}, Y.~G. 1989, Solar System Research, 23, 111

\bibitem[{{Shkuratov} {et~al.}(1984){Shkuratov}, {Akimov}, \&
  {Tishkovets}}]{1984SvAL...10..331S}
{Shkuratov}, Y.~G., {Akimov}, L.~A., \& {Tishkovets}, V.~P. 1984, Soviet
  Astronomy Letters, 10, 331

\bibitem[{{Shkuratov} {et~al.}(1994){Shkuratov}, {Muinonen}, {Bowell}, {Lumme},
  {Peltoniemi}, {Kreslavsky}, {Stankevich}, {Tishkovetz}, {Opanasenko}, \&
  {Melkumova}}]{1994EM&P...65..201S}
{Shkuratov}, Y.~G., {Muinonen}, K., {Bowell}, E., {et~al.} 1994, Earth Moon and
  Planets, 65, 201

\bibitem[{Shkuratov \& Ovcharenko(2002)}]{Shkuratov2002NATO}
Shkuratov, Y.~G. \& Ovcharenko, A.~V. 2002, Experimental Modeling of Opposition
  Effect and Negative Polarization of Regolith-Like Surfaces, ed. G.~Videen \&
  M.~Kocifaj (Dordrecht: Springer Netherlands), 225--238

\bibitem[{{Spadaccia} {et~al.}(2022){Spadaccia}, {Patty}, {Capelo}, {Thomas},
  \& {Pommerol}}]{2022A&A...665A..49S}
{Spadaccia}, S., {Patty}, C.~H.~L., {Capelo}, H.~L., {Thomas}, N., \&
  {Pommerol}, A. 2022, Astronomy and Astrophysics, 665, A49

\bibitem[{{Spadaccia} {et~al.}(2023){Spadaccia}, {Patty}, {Thomas}, \&
  {Pommerol}}]{2023Icar..39615503S}
{Spadaccia}, S., {Patty}, C.~H.~L., {Thomas}, N., \& {Pommerol}, A. 2023,
  Icarus, 396, 115503

\bibitem[{{Spencer} {et~al.}(1989){Spencer}, {Lebofsky}, \&
  {Sykes}}]{1989Icar...78..337S}
{Spencer}, J.~R., {Lebofsky}, L.~A., \& {Sykes}, M.~V. 1989, Icarus, 78, 337

\bibitem[{{Sultana} {et~al.}(2023){Sultana}, {Poch}, {Beck}, {Schmitt},
  {Quirico}, {Spadaccia}, {Patty}, {Pommerol}, {Maturilli}, {Helbert}, \&
  {Alemanno}}]{2023Icar..39515492S}
{Sultana}, R., {Poch}, O., {Beck}, P., {et~al.} 2023, Icarus, 395, 115492

\bibitem[{{Tapping} \& {Reilly}(1986)}]{1986JOSAA...3..610T}
{Tapping}, J. \& {Reilly}, M.~L. 1986, Journal of the Optical Society of
  America A, 3, 610

\bibitem[{Tropf \& Thomas(1997)}]{1997_Tropf_Alumina}
Tropf, W.~J. \& Thomas, M.~E. 1997, in Handbook of Optical Constants of Solids
  Volume III, ed. E.~D. Palik (Burlington: Academic Press), 653--682

\bibitem[{{Tsuda} {et~al.}(2013){Tsuda}, {Yoshikawa}, {Abe}, {Minamino}, \&
  {Nakazawa}}]{2013AcAau..91..356T}
{Tsuda}, Y., {Yoshikawa}, M., {Abe}, M., {Minamino}, H., \& {Nakazawa}, S.
  2013, Acta Astronautica, 91, 356

\bibitem[{{Umow}(1905)}]{Umow1905}
{Umow}, v.~N. 1905, Phusikalische Zeitschrift, 6

\bibitem[{{Vedam} {et~al.}(1975){Vedam}, {Kirk}, \&
  {Achar}}]{1975JSSCh..12..213V}
{Vedam}, K., {Kirk}, J.~L., \& {Achar}, B.~N.~N. 1975, Journal of Solid State
  Chemistry France, 12, 213

\bibitem[{{Veverka} \& {Noland}(1973)}]{1973Icar...19..230V}
{Veverka}, J. \& {Noland}, M. 1973, Icarus, 19, 230

\bibitem[{{Volten} {et~al.}(2001){Volten}, {Mu{\~n}oz}, {Rol}, {de Haan},
  {Vassen}, {Hovenier}, {Muinonen}, \& {Nousiainen}}]{2001JGR...10617375V}
{Volten}, H., {Mu{\~n}oz}, O., {Rol}, E., {et~al.} 2001, Journal of Geophysics
  Research, 106, 17375

\bibitem[{{Watanabe} {et~al.}(2017){Watanabe}, {Tsuda}, {Yoshikawa}, {Tanaka},
  {Saiki}, \& {Nakazawa}}]{2017SSRv..208....3W}
{Watanabe}, S.-i., {Tsuda}, Y., {Yoshikawa}, M., {et~al.} 2017, Space Science
  Reviews, 208, 3

\bibitem[{Weber(2002)}]{2002_Weber}
Weber, M.~J. 2002, Handbook of Optical Materials (CRC Press)

\bibitem[{{Widorn}(1967)}]{1967AnWiD..27..109W}
{Widorn}, T. 1967, Annalen der Universitaets-Sternwarte Wien, Dritter Folge,
  27, 109

\bibitem[{{Wolff}(1975)}]{1975ApOpt..14.1395W}
{Wolff}, M. 1975, Applied Optics, 14, 1395

\bibitem[{{Wolff}(1980)}]{1980Icar...44..780W}
{Wolff}, M. 1980, Icarus, 44, 780

\bibitem[{{Wolff}(1981)}]{1981ApOpt..20.2493W}
{Wolff}, M. 1981, Applied Optics, 20, 2493

\bibitem[{{Worms} {et~al.}(2000){Worms}, {Renard}, {Hadamcik}, {Brun-Huret}, \&
  {Levasseur-Regourd}}]{2000P&SS...48..493W}
{Worms}, J.~C., {Renard}, J.~B., {Hadamcik}, E., {Brun-Huret}, N., \&
  {Levasseur-Regourd}, A.~C. 2000, Planetary Space Science, 48, 493

\bibitem[{{Zellner} {et~al.}(1977{\natexlab{a}}){Zellner}, {Leake}, {Lebertre},
  {Duseaux}, \& {Dollfus}}]{1977LPSC....8.1091Z}
{Zellner}, B., {Leake}, M., {Lebertre}, T., {Duseaux}, M., \& {Dollfus}, A.
  1977{\natexlab{a}}, Lunar and Planetary Science Conference Proceedings, 1,
  1091

\bibitem[{{Zellner} {et~al.}(1977{\natexlab{b}}){Zellner}, {Lebertre}, \&
  {Day}}]{1977LPSC....8.1111Z}
{Zellner}, B., {Lebertre}, T., \& {Day}, K. 1977{\natexlab{b}}, Lunar and
  Planetary Science Conference Proceedings, 1, 1111

\end{thebibliography}


\begin{appendix}
\section{Parameter Space Related to $\amin$}\label{s:amin space}
\begin{figure*}[tb!]
\centering
\includegraphics[width=1\linewidth]{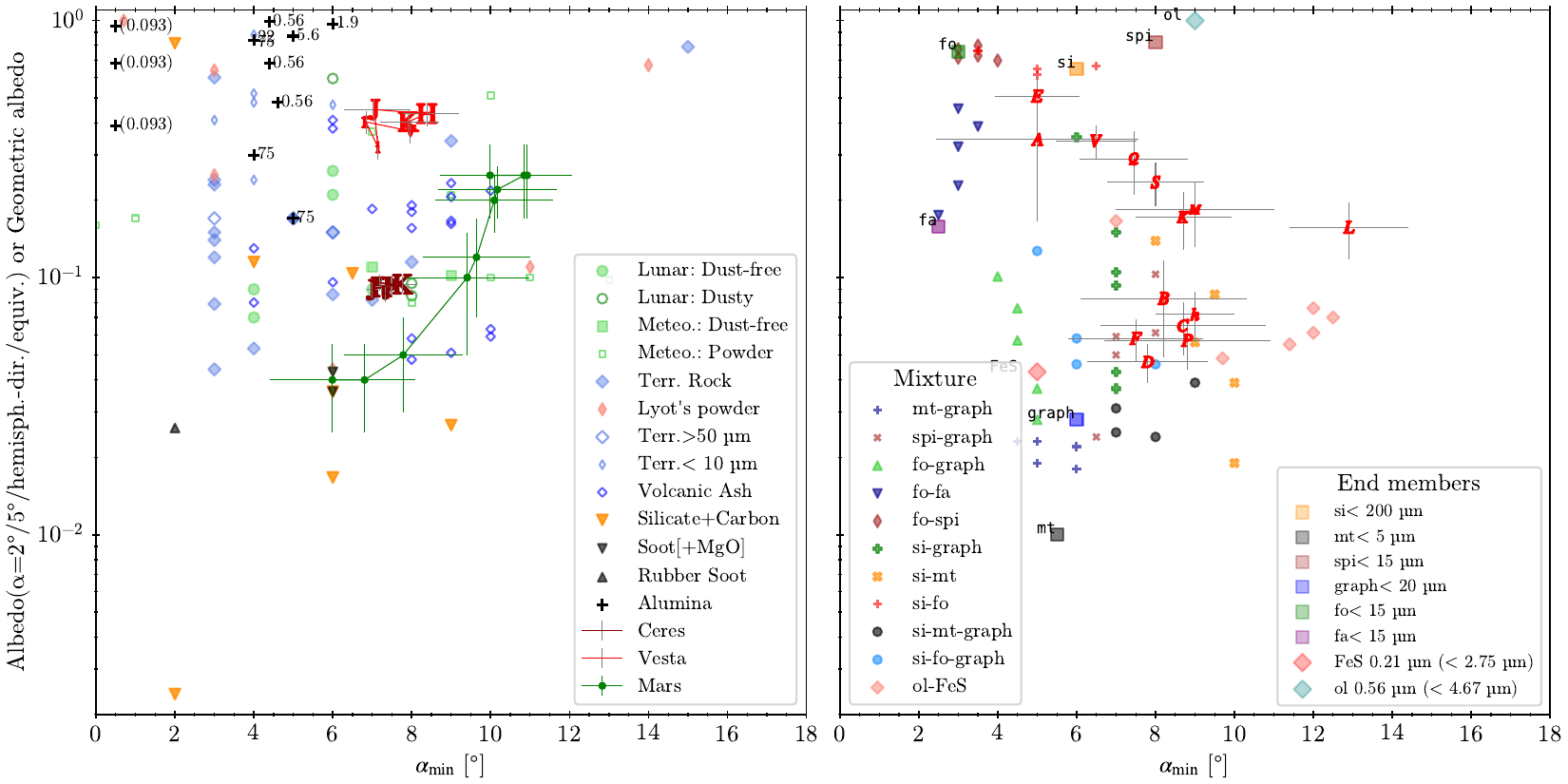}
\caption{Same as Fig. \ref{fig:a0-albedo} but with $\ainv-\amin$. The $\amin$ values are from \cite{2017Icar..284...30B}, while that of Q-type is adopted for (214869) 2007 VA8 (\citealt{gil-hutton-data}; \url{http://gcpsj.sdf-eu.org/catalogo.html}). Ceres and Vesta are shown for comparison (see Fig. \ref{fig:h-albedo}). Data points for Mars are connected in the order of increasing wavelength.}
\label{fig:amin-albedo}
\end{figure*}

\begin{figure*}[tb!]
\centering
\includegraphics[width=1\linewidth]{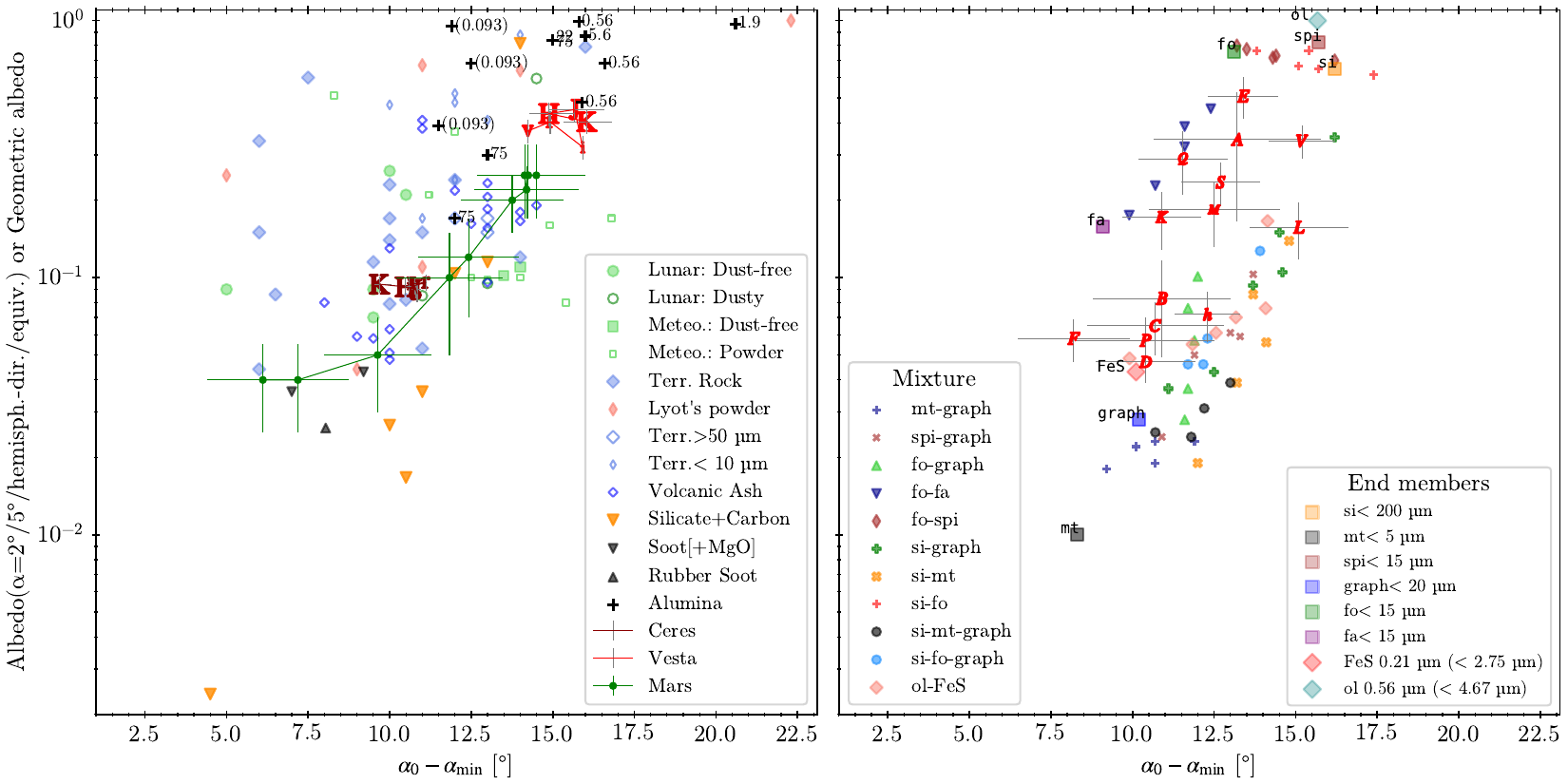}
\caption{Same as Fig. \ref{fig:amin-albedo} but with $\ainv-\amin$.}
\label{fig:a0-amin-albedo}
\end{figure*}

\begin{figure*}[tb!]
\centering
\includegraphics[width=1\linewidth]{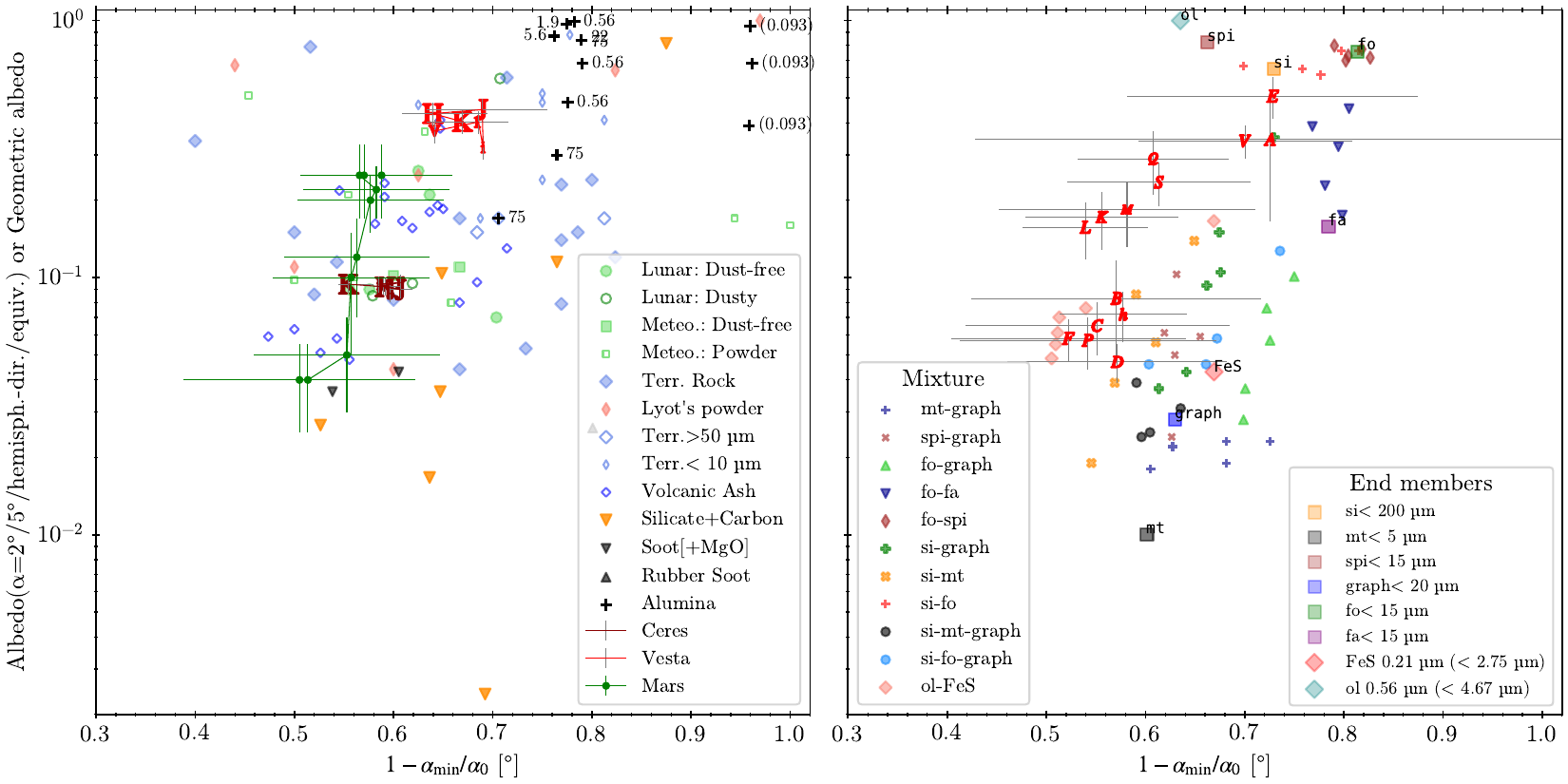}
\caption{Same as Fig. \ref{fig:amin-albedo} but with $1-\amin/\ainv$.}
\label{fig:a0-skew-albedo}
\end{figure*}

Figs. \ref{fig:amin-albedo}, \ref{fig:a0-amin-albedo}, and \ref{fig:a0-skew-albedo} are similar to Fig. \ref{fig:a0-albedo} but with $\amin$ and $\ainv-\amin$, and the skewness $S \coloneqq 1 - \amin/\ainv$ (larger $S$ value may indicate a strong POE) on the abscissa. Interestingly, the mixture samples show a tighter relationship in the latter two parameter spaces. Additionally, asteroids \citep{2017Icar..284...30B} and Mars (Sect. \ref{ss:mars}) are located in very similar locations. Even the L-type that shows a distinctive location in the albedo-$\ainv$ space is now located in the general trend for the latter two diagrams. The asteroids and Mars generally had less skewed NPB than did the laboratory samples (especially $0.5 \lesssim S \lesssim 0.6$ for albedo $\lesssim 30\%$ objects).

\end{appendix}


\end{document}